\documentclass{statsoc}
\usepackage{chngcntr}
\usepackage[a4paper]{geometry}
\usepackage{graphicx}
\usepackage[textwidth=8em,textsize=small]{todonotes}
\usepackage{amsmath}
\usepackage{natbib}
\usepackage{url}
\usepackage{amsbsy}
\usepackage{amssymb}
\usepackage{amsfonts}
\usepackage{mathtools}
\usepackage[ruled,vlined]{algorithm2e}
\usepackage{comment}
\usepackage{multirow}

\newcommand{\ffrac}[2]{\ensuremath{\frac{\displaystyle #1}{\displaystyle #2}}}

\let\oldnl\nl
\newcommand{\nonl}{\renewcommand{\nl}{\let\nl\oldnl}}

\title[Bayesian Design for Spatial Processes]{Bayesian Design with Sampling Windows for Complex Spatial Processes}

\author[K. Buchhorn {\it et al.}]{K. Buchhorn$^{1,2}$, K. Mengersen$^{1,2}$, E. Santos-Fernández$^{1,2}$, E.E. Peterson$^{1,4}$, J.M. McGree$^{1,2,3}$} 
\address{$^{1}$School of Mathematical Sciences. Queensland University of Technology.}
\address{$^{2}$Centre for Data Science (CDS). Queensland University of Technology.}
\address{$^{3}$Australian Institute of Marine Science (AIMS). Crawley, Western Australia.}
\address{$^{4}$EP Consulting. Brisbane, Queensland.}

\begin{document}
\begin{abstract}

Optimal design facilitates intelligent data collection.
In this paper, we introduce a fully Bayesian design approach for spatial processes with complex covariance structures, like those typically exhibited in natural ecosystems.
Coordinate Exchange algorithms are commonly used to find optimal design points.
However, collecting data at specific points is often infeasible in practice.
Currently, there is no provision to allow for flexibility in the choice of design.
We also propose an approach to find Bayesian sampling windows, rather than points, via Gaussian process emulation to identify regions of high design efficiency across a multi-dimensional space.
These developments are motivated by two ecological case studies: monitoring water temperature in a river network system in the northwestern United States and monitoring submerged coral reefs off the north-west coast of Australia.

\end{abstract}

\section{Introduction}

Sampling designs that optimise one or more specified utility functions are valuable in many applied contexts, since they facilitate the intelligent collection of data. 
That is, an optimal design results in greater information efficiency, reduced sampling cost and improved estimation. 
However, such schemes remain challenging to develop for spatial processes with complex covariance structures, such as those found in natural ecosystems. 
Moreover, sampling at the exact design point can be difficult in practice within these systems due to practical challenges associated with the collection of samples and accessibility of the sites.
In this case, a more flexible sampling scheme that identifies windows with near-optimal design efficiency can be highly useful.

Data collected by environmental monitoring programs are essential for understanding patterns, trends and vulnerabilities in ecological and environmental systems. 
Here, we focus on river network and coral reef systems, and develop optimal sampling regions given non-standard spatial relationships.
The covariance across locations on a branching river is particularly complex, since the network topology is embedded in a 2D terrestrial landscape, with directional water flow, and disparities between flow volume.
Furthermore, monitoring programs are notoriously expensive in terms of monetary, human and technological resources \citep{Lind2010the,Roelf2010calibration}. 
Balancing short- and long-term requirements, as well as practical constraints of data collection can add further complication to the design process \citep{kang2016bayesian}. 
Our aim is to develop a more flexible design approach within a Bayesian framework while achieving two goals: 1) the determination of a set of optimal sampling locations based on a discrete set of available locations, and 2) the formation of sampling windows within a specified region where difficulties due to access or otherwise may be encountered. 


Experimental design problems are commonly viewed as optimisation problems.
In the classical framework, optimal experimental designs are often derived using utilities based on the expected Fisher information matrix \citep[e.g.][]{atkinson1992optimum}. 
Such designs have been shown to depend on assumed values for model parameters, so approaches have been proposed to remove some of this dependence. 
For example, \cite{pronzato1985robust} optimised a design criterion over a prior probability distribution for the model parameters; the so-called `pseudo-Bayesian' design approach.
While pseudo-Bayesian designs are more robust for the choice of parameter values, the prior information is not utilised in subsequent analysis (i.e. computation of the utility). 
A fully Bayesian framework provides a unified approach for incorporating prior information and rigorously handling uncertainty about the model parameters and the statistical model when forming a design and analysing the resulting data (see \cite{chaloner1995bayesian, ryan2016review}). 

Various approaches to finding sampling intervals, so-called windows, are presented in the statistical literature relating to pharmacokinetics (e.g. \cite{bogacka2008d, foo2012general,mcgree2012sequential}). 
These intervals provide flexibility in the timing of sample collection, while 
assuring a high level of information efficiency.
An approach for stratified random sampling in a geostatistical simulation study by \cite{lin2011modeling} uses Latin Hypercube sampling to find the optimal sub-sampling of pre-specified regions (for a given spatial stratification).
However, to our knowledge no approaches have been developed to form sampling windows for spatial processes in the statistical literature.

We present a Bayesian approach, using a Gaussian process emulator to approximate the utility surface in high-dimensions.
Our method builds on the approach to stochastic optimisation employed in the Approximate Coordinate Exchange (ACE) \citep{overstall2017bayesian} algorithm. 
We also demonstrate how to derive design efficiency contours using sampling windows, which can then be used to guide highly efficient and flexible sampling of complex spatial processes.
The new approach is applied to substantive real-world case studies, described below.


\subsection{Motivational Case Studies}


Example 1: Water temperature is considered a ``master" variable in river ecosystems because it affects the metabolic rates of aquatic organisms, life history events related to reproduction, hatching and maturation, and restricts the distribution and abundance of ectothermic species \citep{isaak2017norwest}. 
Water temperature is often monitored to understand the impacts of climate change and land management on thermal habitats \citep{isaak2016slow, jackson2018spatio} and is also used as the basis for regulatory actions \citep{todd2008development}. 
The recent proliferation of inexpensive \emph{in-situ} sensors has dramatically increased the potential for monitoring and data-enabled management decisions \citep{isaak2017norwest}. 
Nevertheless, comprehensive monitoring on river networks using in-situ sensors remains costly and complex in terms of balancing short- and long-term priorities. 
Optimal sensor placement must be assessed when sensors are initially deployed, new sensors are added to an established monitoring program or the number of in-situ sensors must be reduced. 
In addition, field conditions can be unpredictable; some sites are inaccessible due to steep terrain, a lack of network reception for data streaming, or private landholders refusing access.
This motivates two considerations: the design of a global optimal sampling scheme across the river network, and a more targeted approach to identify finer-scale sampling regions that provide near-optimal design efficiency.

In river networks, spatial relationships are characterised in part by their topology (e.g. branching network structure, connectivity).
There have been numerous advances in statistical modelling used to describe spatial covariance in data collected on branching river networks based on distance travelled along the river, directionality in water flow and differences in flow volume \citep{ver2006spatial, ver2010moving, peterson2013modelling, santos2021bayesian}. 
The ability to model this spatial component and more accurately predict conditions throughout a river network often provides a better understanding of habitat suitability \citep{isaak2016slow} and leads to more effective management decisions \citep{fig2021guiding,sharma2021dendritic}. 
Frequentist sampling methods for river networks have been developed \citep{som2014spatial,wang2020designing}, as well as pseudo-Bayesian approaches \citep{falk2014sampling, pearse2020ssndesign}. 
However, a fully Bayesian design framework for river networks is currently unexplored. 



Example 2: Coral reefs are inextricably linked by the symbiotic relationships formed between corals and the organisms around them \citep{richmond1993coral}. 
In particular, deep reefs have the capacity to act as refugia and to assist in the recovery of more vulnerable shallow water reefs following a disturbance. 
However, major challenges remain in understanding the extent of and managing the vast marine estate for Australia, as well as for many other maritime nations.
Insights into remote reefs are beginning to emerge due to data captured using new technologies such as remote and autonomous underwater vehicles. 
With limited resources to invest in research and monitoring, there is a need to optimise in-field activities to collect data that will most efficiently achieve research and management goals.


Hard coral cover is an indicator commonly used to infer the health and condition of a coral reef \citep{ltmp2022}.
In this example, we consider deep reefs, or mesophotic coral ecosystems, with data collected between 12 to 50 metres in depth.
The extent to which important ecological processes change along depth gradients in a reef is not well understood and is a critical knowledge gap for developing future reef policies and management practices \citep{kang2016bayesian}.
Monitoring coral reef environments often relies on a series of images taken underwater, along a transect line. 
As such, collecting data from submerged shoals can be costly given the specialised skills of those undertaking the monitoring and the cost of equipment and running a ship.
Motivating our second design scenario, we consider the optimal placement of transect lines across a submerged shoal taking into account the spatial nature of the process. 
In practice, underwater sampling along these lines can be challenging and unpredictable due to water currents. 
Thus, a more practical solution would be to define sampling regions, as proposed here, offering more flexibility and assurance in sampling.


The following section provides details about the general statistical modelling used in both case studies, then outlines specific details in each case. 
In Section 3, the Bayesian optimal design approach is described. 
Our proposed approach to finding sampling windows is given in Section 4, along with the concept of design efficiency contours. 
Finally, Section 5 illustrates the results of these two examples and demonstrates the scope of applicability of Bayesian design and sampling windows for spatial processes.

\section{Model}
\subsection{Bayesian Spatial Generalised Linear Models}

We start with a general description of spatial linear models and then extend this to models for data collected on river networks and coral reefs. 
Let $\textbf{d} = (d^1, ..., d^\gamma)$ denote a design, that is, a set of locations in the physical or parameter space.
Consider a linear mixed model with $n \times 1$ response $\textbf{Y}$, and the $n \times p$ design matrix $\textbf{X}$ of explanatory variables spatially indexed at locations, $\textbf{d}$,
\begin{equation}
    \textbf{Y} = \textbf{X} \boldsymbol{\beta} + \textbf{z} + \boldsymbol{\epsilon}, \label{linearmodel}
\end{equation}

\noindent with spatially correlated random effects $\textbf{z} \sim \mathcal{N}(\bf{0}, \boldsymbol{\Sigma_z})$ and independent residuals $\boldsymbol{\epsilon} \stackrel{iid}{\sim} \mathcal{N}(0, \sigma_0^2)$ yielding a covariance of $\boldsymbol{\Sigma} = \mbox{Cov}[\textbf{Y}] = \boldsymbol{\Sigma}_z + \sigma_0^2 \textbf{I}$ \noindent where \textbf{I} is the $n \times n$ identity matrix and $\sigma_0^2$ is the nugget effect. Typically, we express $\boldsymbol{\Sigma_z}$ as a function based on distance $h$ between locations $\textbf{d}$, and covariance parameters, $\boldsymbol{\theta}_z$. We define the set of all parameters as $\boldsymbol{\theta} = (\boldsymbol{\beta}, \boldsymbol{\theta}_z)'$. 

There are many covariance models that describe how the dependence between observed data at two spatial locations decays with distance. A common example is the exponential function,
\begin{equation}
    C(h|\boldsymbol{\theta}_z) = \sigma_1^2 \mbox{exp}\Big(\frac{-3h}{\alpha}\Big),
\label{eq:exp}
\end{equation}

\noindent where the partial sill, $\sigma_1^2$, represents the magnitude of variance and the range parameter, $\alpha$, describes how fast the covariance decays with an appropriate measure of distance between the locations, $h$.

The generalised linear modelling (GLM) framework provides a unified approach to analyse data for which some function of the mean response (link function) may vary linearly with the predictors rather than the mean response itself. 
We denote the linear predictor by $\textbf{X} \boldsymbol{\beta} = \boldsymbol{\eta} = \beta_0 + \sum\limits^{p}_{j=1}\textbf{X}_{j} \beta_j$. 
A link function $g(\cdot)$ relates the linear predictor to the mean of the outcome variable $g(\mathbb{E}(\textbf{Y}|\textbf{X})) = \boldsymbol{\eta}$. 
In a GLM, $\textbf{Y}$ is assumed to be generated from a distribution in the exponential family, where the probability density (or mass) function is written as,
\begin{equation*}
    p(y_i) = \mbox{exp}\Bigg(\frac{y_i \psi_i - b(\psi_i)}{a_i(\phi)} + c(y_i, \phi) \Bigg),
\end{equation*}
where for observation $i$, $\psi_i$ is the location parameter, $\phi$ is the scale parameter and $a_i(\cdot)$, $b(\cdot)$ and $c(\cdot,\cdot)$ are known functions.\\

\noindent \emph{Example 1: Model for River Networks}

\noindent Consider data collected on a river network exhibiting complex patterns of spatial dependence due to multi-scale processes occurring within and between the aquatic and terrestrial environments \citep{peterson2013modelling}. 
Two points $d^k$ and $d^s$ on the river network are said to be \emph{flow-connected} if water flows from an upstream site to a downstream site. They are \emph{flow-unconnected} if they reside on the same river network and share a common junction downstream, but do not share flow. 
Following \cite{ver2010moving}, we estimate autocorrelation between points $d^k$ and $d^s$ by constructing random variables as the integration of a moving-average function over a white-noise random process.
The tails of these moving-average functions are unilateral and spatial autocorrelation only occurs when the tails of the functions overlap.
If the tail points upstream, it is referred to as a \emph{tail-up} model and by construction, these models restrict correlation to flow-connected locations. Stream networks are dendritic and so weights are used to proportionally allocate (i.e. split) the tail-up moving average function at river junctions. 
Tail-up models are particularly useful for modelling organisms or materials that flow passively downstream (e.g. water pollutants).
Conversely, when the tail of the function points downstream it is referred to as a \emph{tail-down} model. These models allow spatial correlation between both flow-connected and flow unconnected locations and may be more suitable for modelling organisms such as fish that actively move both up and downstream \citep{peterson2003upstream}. 

A mixed modelling approach is often used for modelling river network data since it allows for more complex patterns of spatial dependence to be described within a single model \citep{peterson2010mixed}. Variance components are used to expand the spatially dependent random variable $\textbf{z}$ into several zero-mean random variables, 
\begin{equation}
\textbf{Y} = \textbf{X} \boldsymbol{\beta} + \textbf{z}_{_{EUC}} + \textbf{z}_{_{TU}} + \textbf{z}_{_{TD}} + \boldsymbol{\epsilon},
\label{eq:mixed}
\end{equation}

\noindent with correlation structure described using Euclidean (EUC), tail-up (TU) and tail-down (TD) covariance models. 
As the notation suggests, the general covariance matrix can be formulated by a combination of such covariance models,
\begin{align}
\boldsymbol\Sigma &= \mbox{Cov}(\textbf{z}_{_{EUC}}) + \mbox{Cov}(\textbf{z}_{_{TU}}) + \mbox{Cov}(\textbf{z}_{_{TD}}) + \mbox{Cov}(\boldsymbol{\epsilon}) \nonumber\\
&= \textbf{C}_{_{EUC}}(h_{_{EUC}}|\boldsymbol{\theta}_{\textbf{z}_{EUC}}) +  \textbf{C}_{_{TU}}(h_{_{H}}|\boldsymbol{\theta}_{\textbf{z}_{TU}}) +  \textbf{C}_{_{TD}}(h_{_{H}}|\boldsymbol{\theta}_{\textbf{z}_{TD}}) + \sigma_0^2 \textbf{I}, \label{eqsigma}
\end{align}
\noindent where $h_{_{EUC}}$ is the Euclidean distance and $h_{_{H}}$ represents the hydrological distance, that is, distance \emph{along} the river. 
Note that, distance considered can depend on flow-connectivity of sites and the appropriate covariance function but for brevity we denote this as $h_{_{H}}$, see \cite{santos2021bayesian} for further details.
See Appendix A.1 for the specification of the covariance matrices $\textbf{C}_{_{EUC}}$, $\textbf{C}_{_{TU}}$ and $\textbf{C}_{_{TD}}$, as well as distance $h_{_{H}}$.

The generalised linear mixed model for river networks we consider is of the form,
\begin{align}
   \textbf{Y} &\sim \mathcal{N}(\boldsymbol{\mu}, \boldsymbol\Sigma)\label{eqlinear}\\
   p(\textbf{y}|\boldsymbol{\theta}, \textbf{d}) = (2\pi)^{-\frac{k}{2}}\mbox{det}(\boldsymbol\Sigma)^{-\frac{1}{2}}&\mbox{exp}\big(-\frac{1}{2}(\textbf{y}-\boldsymbol{\mu})^T\boldsymbol\Sigma^{-1}(\textbf{y}-\boldsymbol{\mu})\big),\nonumber
\end{align}

\noindent with link function as the identity $g(\boldsymbol{\mu}) =\textbf{X}\boldsymbol{\beta}$ and $\boldsymbol\Sigma$ as defined by Equation \eqref{eqsigma} at a collection of locations $\textbf{d}$ on the network.

\vspace{0.5cm}
\noindent \emph{Example 2: Coral Reef Modelling}

\begin{figure}
\includegraphics[width=0.9\textwidth]{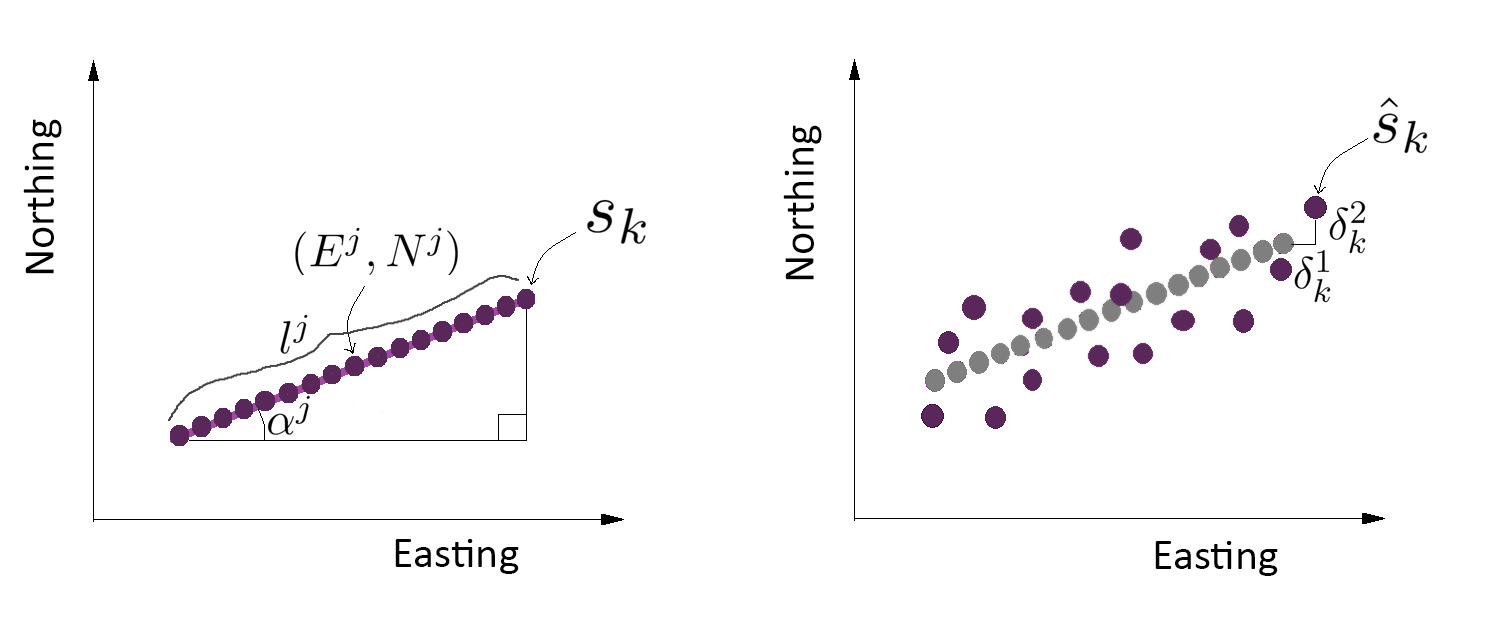}
\caption{Data is sampled at each geographic point $s_k$ along the transect defined by $d^j = (E^j, N^j, \alpha^j, l^j)$ (left). A transect region with radius $r^j$ is shown, where data is collected at the stochastic sample locations $\hat{s}_k$ within the region (right).}
\label{fig:transect}
\end{figure}

\noindent Estimates of hard coral are generally obtained using a number of images, $N$, of the reef floor at locations $s_k$ collected along a transect. 
In order to define transect placement, design parameters are introduced: the midpoint of a transect using Easting and Northing coordinates, $E^j$ and $N^j$, the angle of the transect, $\alpha^j$ in degrees, and the length of the transect, $l^j$ in meters (m).
Transects can therefore be fully specified by design parameters $d^j = (E^j, N^j, \alpha^j, l^j)$ for transect $j$ (see Figure \ref{fig:transect}).
Let $s_k = (e_k, n_k)$ denote the geographic coordinates of image $k$ specified by a collection of transects $\textbf{d}$.
Another parameter, $r^j$, is introduced to create a transect \emph{region}. For radius $r^j > 0$, points disperse across the geographic plane to $\hat{s}_k = (\hat e_k, \hat n_k) = (e_k + \delta_{k}^1, n_k+ \delta_{k}^2)$, with $\delta_{k}^1,\delta_{k}^2 \sim \mbox{Unif}(-r^j, r^j)$. 

For each image a number, $n_k$, of randomly selected points are placed on image, $k$, and classified as hard coral, or not. 
The number of points within an image that contain hard coral, $y_k$, are assumed to have a binomial distribution 
\begin{align}
   \textbf{Y} &\sim \mbox{Binomial}(\textbf{q},\textbf{n}) \nonumber\\
   p(\textbf{y}|\boldsymbol{\theta}, \textbf{d}, \textbf{z}) &= \prod^N_{k=1} { n_k \choose y_k}  q_k^{y_k} (1-q_k)^{n_k-y_k}.
\label{eqn:logit}
\end{align}
where $\textbf{q} = \mathbb{E}(\textbf{Y}|\textbf{X}, \boldsymbol{\theta}, \textbf{z})$, with spatially correlated random effects $\textbf{z} \sim \mathcal{N}(\bf{0}, \boldsymbol{\Sigma_z} + \sigma_0^2 \textbf{I})$ where the matrix $\boldsymbol{\Sigma_z}$ has entries of the form in Equation \eqref{eq:exp} or some other covariance matrix. 
The expected value of coral cover $q_k$ on image $k$ leads to the specification of a logistic regression model with link function denoted by $g(\textbf{q}) = \mbox{log}(\textbf{q}/(1-\textbf{q}))$, so that $\textbf{q} = \mbox{logit}^{-1}(\boldsymbol{\eta} + \textbf{z}) = \mbox{exp}(\boldsymbol{\eta} + \textbf{z})/(1 + \mbox{exp}(\boldsymbol{\eta} + \textbf{z}))$ with $\boldsymbol{\eta} = \textbf{X} \boldsymbol{\beta}$. 


\subsection{Bayesian Framework}

Let $p(\boldsymbol{\theta})$ denote the prior distribution on $\boldsymbol{\theta}$, and consider the joint posterior distribution,
\begin{equation}
p(\boldsymbol{\theta}, \textbf{z}| \textbf{y}, \textbf{d}) = \frac{p(\textbf{y} | \boldsymbol{\beta}, \textbf{d}, \textbf{z}) p(\boldsymbol{\theta} ) p(\textbf{z}|\boldsymbol{\theta}_z) }{p(\textbf{y}|\textbf{d})} \propto p(\textbf{y} | \boldsymbol{\beta}, \textbf{d}, \textbf{z}) p(\boldsymbol{\theta} ) p(\textbf{z}|\boldsymbol{\theta}_z),
\label{eqn:joint}
\end{equation}

\noindent where $p(\textbf{y}|\textbf{d}) = \int_{\theta} p(\textbf{y}, \boldsymbol{\theta})\ d\boldsymbol{\theta}= \int_{\theta} p(\textbf{y} | \boldsymbol{\theta}, \textbf{d} ) p(\boldsymbol{\theta})\ d\boldsymbol{\theta}$ is the marginal distribution of $\textbf{y}$. 
Unless the likelihood and spatial random effect are analytically tractable, that is, $\int_{\textbf{z}} p(\textbf{y} | \boldsymbol{\beta}, \textbf{d}, \textbf{z})p(\textbf{z}|\boldsymbol{\theta}_z)\ d\textbf{z} = p(\textbf{y}|\boldsymbol{\theta}, \textbf{d})$ has a closed form solution, the spatial random effect must be integrated out numerically. 
We describe the approach taken in the next section.

\section{Bayesian Optimal Design for Spatial Models}

Optimal design problems are commonly viewed as an optimisation framework with respect to an experimental aim. A utility function is defined to quantify the worth of a design, which may be to control systematic error (bias), reduce random variation, increase the precision of parameter estimates, make predictions about future observations or discriminate between competing models. 
Formally, we denote the utility function as $u(\textbf{d}, \boldsymbol{\theta}, \textbf{y})$ which depends on model parameters $\boldsymbol{\theta}\sim p(\boldsymbol{\theta})$ and the data we might observe under that statistical model $\textbf{y} \sim p(\textbf{y} | \boldsymbol{\theta}, \textbf{d})$ at design points $\textbf{d}$ from the design space $\mathcal{D}$. 
We establish the notation for a design as follows, $\textbf{d}_i = (d_i^1, ..., d_i^\gamma)$, and use $d_i^j$ to refer to the $j$th coordinate of design $\textbf{d}_i$ throughout the paper.

The aim is to maximise the expected utility, $U(\textbf{d}) = \mathbb{E}[u(\textbf{d}, \boldsymbol{\theta}, \textbf{y})]$ over the entire parameter space $\Theta$ and all conceivable data sets $\mathcal{Y}$. The Bayesian optimal design $\textbf{d}_*$ can therefore be expressed as,
\begin{align}
\textbf{d}_* &= \mbox{arg max}_{\textbf{d} \in \mathcal{D}} \mathbb{E}[u(\textbf{d}, \boldsymbol{\theta}, \textbf{y})] \nonumber \\
&= \mbox{arg max}_{\textbf{d} \in \mathcal{D}} \int_{\textbf{y} \in \mathcal{Y}} \int_{\boldsymbol{\theta} \in \Theta} u(\textbf{d}, \boldsymbol{\theta}, \textbf{y})\ p(\boldsymbol{\theta}, \textbf{y}| \textbf{d}) \ d\boldsymbol{\theta}\ d\textbf{y} \label{eq2} \\
&= \mbox{arg max}_{\textbf{d} \in \mathcal{D}} \int_{\textbf{y} \in \mathcal{Y}} \int_{\boldsymbol{\theta} \in \Theta} u(\textbf{d}, \boldsymbol{\theta}, \textbf{y})\ p(\textbf{y} | \boldsymbol{\theta}, \textbf{d})\ p(\boldsymbol{\theta})\ d\boldsymbol{\theta}\ d\textbf{y}. \nonumber 
\end{align}

Unfortunately, the integral in Equation \eqref{eq2} is analytically intractable for most applications, and indeed the linear mixed-model in Equation \eqref{linearmodel}. In practice, a commonly used method is Monte Carlo integration, where the expected utility can be approximated by the empirical mean, that is,

\begin{equation}
\hat U(\textbf{d}) = \frac {1}{M} \sum^{M}_{m=1} u(\textbf{d}, \boldsymbol{\theta}^{(m)}, \textbf{y}^{(m)})
\label{eq:monte}
\end{equation}

\noindent with draws from the prior $\boldsymbol{\theta}^{(m)} \sim p(\boldsymbol{\theta})$ and then the likelihood $\textbf{y}^{(m)} \sim p(\textbf{y} | \boldsymbol{\theta}^{(m)}, \textbf{d})$. 
In order to accurately estimate $\hat U(\textbf{d})$, $M$ needs to be large. 
In some cases, the median $\tilde U(\textbf{d}) =  \mbox{median}(u(\textbf{d}, \boldsymbol{\theta}^{(m)}, \textbf{y}^{(m)}))$ may be preferred to the arithmetic mean in Equation \eqref{eq:monte}, since it is less sensitive to tail behaviour when the target distribution is asymmetric. 
We explore this comparison further in the following sections.

A commonly used utility function is the Kullback-Liebler (KL) divergence \citep{kullback1951information} of the posterior from the prior distribution,
\begin{align}
\label{eqn:kld}
u(\textbf{y}, \textbf{d}) &= D_{KL}[p(\boldsymbol{\theta} | \textbf{y}, \textbf{d})||p(\boldsymbol{\theta})] \nonumber \\
&=  \int p(\boldsymbol{\theta} | \textbf{y}, \textbf{d}) \log \frac{p(\boldsymbol{\theta} | \textbf{y}, \textbf{d})}{p(\boldsymbol{\theta})} \ \mbox{d}\theta.
\end{align}

\noindent The intuition behind Equation \eqref{eqn:kld} is that a large $D_{KL}$ divergence from posterior to prior implies that the data in $\textbf{y}$ reduces entropy (randomness or uncertainty) in $\theta$ and hence the data are more informative at design points $\textbf{d}$. As suggested by \cite{lindley1956measure}, this utility can be used in the interest of maximising the expected information gain on the model parameters (or functions of these) when performing experiments at design points $\textbf{d}$. When both the prior and the posterior distributions follow multivariate normal distributions, that is, $\boldsymbol{\theta} \sim  \mathcal{N}(\boldsymbol{\mu}_{0},\boldsymbol{ \Sigma_{0}})$ and $\boldsymbol{\theta} | \textbf{y}, \textbf{d} \sim  \mathcal{N}(\boldsymbol{\mu}_{1}, \boldsymbol{\Sigma_{1}})$, respectively, then the KL divergence becomes
\begin{equation*}
u(\textbf{y}, \textbf{d}) = \frac{1}{2} \Big[(\boldsymbol{\mu}_{0} - \boldsymbol{\mu}_{1})^T \boldsymbol{\Sigma_{0}}^{-1}(\boldsymbol{\mu}_{0} - \boldsymbol{\mu}_{1}) + \text{tr}(\boldsymbol{\Sigma_{0}}^{-1}\boldsymbol{\Sigma_{1}}) - \text{ln}\frac{|\boldsymbol{\Sigma_{1}}|}{|\boldsymbol{\Sigma_{0}}|} - {\kappa} \Big],
\end{equation*}

\noindent where $\boldsymbol{\mu}_{0},\ \boldsymbol{\mu}_{1} \in \mathbb{R}^{\kappa}$ and $\boldsymbol{\Sigma_{0}},\ \boldsymbol{\Sigma_{1}} \in \mathbb{R}^{{\kappa} \times {\kappa}}$ \citep{duchi2007derivations}. 

Note that in Bayesian design, the utility is always a function of the posterior distribution, which generally means that inference is performed many thousands of times since the posterior $p(\boldsymbol{\theta} | \textbf{y}, \textbf{d})$ must be evaluated for each future data set $\{\boldsymbol{\theta}^{(m)}, \textbf{y}^{(m)}\}$ that is drawn from the joint distribution $p(\boldsymbol{\theta},\textbf{y}| \textbf{d})$. 
As a computationally efficient approximation to the posterior distribution, we employ the Laplace approximation which has the following form:
\begin{equation}
    \boldsymbol{\theta} | \textbf{y}, \textbf{d} \sim  \mathcal{N}(\boldsymbol{\theta}, \textbf{H}(\boldsymbol{\theta}^{*})^{-1}),
\end{equation}
where $\boldsymbol{\theta}^* = \mbox{arg max}_{\boldsymbol{\theta} \in \Theta}\{\mbox{log } p(\textbf{y}|\boldsymbol{\theta}, \textbf{d}) + \mbox{log }p(\boldsymbol{\theta})\}$ and $\textbf{H}(\boldsymbol{\theta}^{*})$ is the Hessian matrix defined as:
\begin{equation*}
    \textbf{H}(\boldsymbol{\theta}^{*}) = \left.\ffrac{-\partial^2 \{\mbox{log } p(\textbf{y}|\boldsymbol{\theta}, \textbf{d}) + \mbox{log }p(\boldsymbol{\theta})\} }{\partial\boldsymbol{\theta}\partial\boldsymbol{\theta}^{'}}\right\rvert_{\boldsymbol{\theta}=\boldsymbol{\theta}^{*}}.
\end{equation*}

This approximation requires evaluation of the full data likelihood. 
Thus, the full data likelihood used in forming the Laplace approximation is replaced with a numerical approximation to the posterior for $\boldsymbol{\theta}$,
\begin{equation}
p(\boldsymbol{\theta} | \textbf{y}, \textbf{d}) \propto p(\boldsymbol{\theta} ) \int_{\textbf{z}} p(\textbf{y} | \boldsymbol{\beta}, \textbf{d}, \textbf{z})  p(\textbf{z}|\boldsymbol{\theta}_z)\ d\textbf{z}.
\label{eqn:fulllikelihood}
\end{equation}

Finally, an approach for the maximisation required in finding the Bayesian optimal design in Equation \eqref{eq2} needs to be specified. 
It is common to use a Coordinate Exchange (CE) algorithm to search a discrete space, by exchanging coordinates in the design, $\textbf{d}$, with other points in the design space, in a sequential manner. 
Formally, denote a proposal design by $\textbf{d}^{j\curvearrowright} = (d^1, ..., d^{j-1}, d^{\curvearrowright}, d^{j+1}, ..., d^\gamma)$, obtained by exchanging the $j$th design coordinate for some other design point $d^{\curvearrowright} \in \mathcal{D} \setminus \textbf{d}$.
Given the stochastic nature of the utility approximation, illustrated by draws of the prior and likelihood in Equation \eqref{eq2}, the search algorithm needs to accommodate different realisations of $\tilde{U}(\textbf{d})$. 
For this, an acceptance probability $p^*$ can be used to compare a proposed design $\textbf{d}^{j\curvearrowright}$ to the current best design $\textbf{d}$. 
The Approximate Coordinate Exchange (ACE) algorithm \citep{overstall2017bayesian} is a commonly used stochastic CE algorithm with acceptance probability defined by

\begin{equation*}
p^* = 1- T_{2B -2}\Big(- \dfrac{B\hat{U}(\textbf{d}^{j\curvearrowright}) - B\hat{U}(\textbf{d})}{\sqrt{2B\hat{v_b}}}\Big),
\end{equation*}
and 
\begin{equation*}
\hat{v_b} = \dfrac{\sum^B_{l=1} \big[ u(\textbf{d}^{j\curvearrowright}, \boldsymbol{\theta}^{(l)}, \textbf{y}^{(l)}) - \hat{U}(\textbf{d}^{j\curvearrowright})\big]^2 + \sum^B_{l=1} \big[ u(\textbf{d}, \boldsymbol{\theta}^{(l)}, \textbf{y}^{(l)}) - \hat{U}(\textbf{d})\big]^2}{2B -2}
\end{equation*}
with large $B$ and random draws $\{\boldsymbol{\theta}^{(l)}, \textbf{y}^{(l)}\}$ from $p(\boldsymbol{\theta}, \textbf{y} | \textbf{d})$. However, this acceptance criteria relies on the assumption of a normally distributed utility (at least approximately).

We propose an alternative version of the stochastic Coordinate Exchange algorithm (see Appendix Algorithm \ref{stochastic_search}) with a non-parametric acceptance criterion defined as $$p_W^* = 1- W(U(\textbf{d}^{j\curvearrowright}), U(\textbf{d}))$$ where $W$ is the p-value of the one-sided Wilcoxon rank sum test (equivalent to the Mann-Whitney test). 
In this case, the null hypothesis is that the distributions of the proposed design utility, $U(\textbf{d}^{j\curvearrowright})$, and the current best design utility, $U(\textbf{d})$, differ by a location shift of $\mu=0$ and the alternative hypothesis is that $\mu > 0$. The `one-sided' specifies that $U(\textbf{d}^{j\curvearrowright})$ is shifted to the right of $U(\textbf{d})$. We explore the proposed non-parametric stochastic Coordinate Exchange algorithm in comparison to the ACE in the Appendix, both with median and mean central tendency measure.

\section{Sampling Windows}

As mentioned previously, optimal design methods have been used to determine sampling windows for experiments in pharmacokinetics \citep{foo2012general, bogacka2008d}.
For spatial processes, sampling windows would allow the experimenter some flexibility in the sampling location, while preserving a minimum required efficiency for the design utility. 
In order to define sampling windows, we define the utility in a continuous domain.
Consider the design space $\mathcal{D} \subseteq \mathbb{R}^q$ across $q$ windows.
We first introduce Gaussian processes (GP) and illustrate their flexibility and capability as a multi-dimensional utility smoothing technique. 
Based on the smoothing GPs provide, we then show how design efficiency contours can be derived.

GPs generalise the concept of Gaussian distributions over discrete random variables to the function space. 
In the Bayesian context, GPs can be seen as a prior over the function space, where inference takes place. Let $f(\textbf{x})$ denote the target value of interest where $\textbf{x} \in \mathbb{R}^{q}$ is a $q$-dimensional random vector containing predictor values such as spatial locations. 
We write $\mathcal{GP}(m(\textbf{x}), k(\textbf{x}, \textbf{x}^{'}))$ to denote a GP characterised by a mean function $m(\textbf{x})$ and a covariance function $k(\textbf{x}, \textbf{x}^{'})$. 
Consider a GP model with kernel,
\begin{equation}
k(\textbf{x}_p, \textbf{x}_s) = \sum\limits^{q}_{j=1} \exp(-\zeta_j * \mbox{dist}_j(\textbf{x}_p, \textbf{x}_s))
\label{eq:kernel}
\end{equation}
with hyperparameters $\boldsymbol\zeta$ and distance measure defined as $\mbox{dist}_j(\textbf{x}_p, \textbf{x}_s) = |x_{p}^j - x_{s}^j|$, for $j=\{1,...,q\}$. Using a finite collection of inputs $\{\textbf{x}_1,...,\textbf{x}_n\}$ and the function above, we construct the kernel matrix $\textbf{K}(\cdot)$ over the parameter space.

Details of the proposed Sampling Windows Algorithm are presented in Algorithm \ref{gp}, using design notation. We begin by defining a design space for windows (line 2), this may differ from the design space searched in the discrete stochastic CE algorithm.
We fit a GP (line 4) to the criterion of interest, here, the median utility $\tilde U(\textbf{d}_i)$ computed at each input $\textbf{d}_i$ (line 3), assuming a zero-mean GP prior, $f \sim \mathcal{GP}(\textbf{0}, \textbf{K}(\textbf{D},\textbf{D}) + \zeta_0 \textbf{I})$. 
Inclusion of the nugget $\zeta_0 > 0$ ensures that every $f$ has some magnitude of variance with itself only (independent noise), so that the Monte Carlo approximations of the median utility surface will be smoothed, not interpolated. 
Hyperparameters $\boldsymbol\zeta = (\zeta_0,...,\zeta_{\kappa})$ are determined (line 5) by minimising cross-validation error of the GP, 
\begin{align}
\hat{\boldsymbol{\zeta}} &= \underset{\boldsymbol{\zeta}}{\mathrm{arg\ min}}\ CV(\boldsymbol{\zeta}|f,\textbf{D}) \text{, with }\\
CV(\boldsymbol{\zeta}|f,\textbf{D}) &= \ffrac{1}{\gamma_1} \sum\limits^{\gamma_1}_{i=1} \Bigg(\ffrac{\tilde{U}(\textbf{d}_i|\textbf{d}_c) - f(\textbf{d}_i)}{1 - S_{ii}}\Bigg)^2
\label{eqn:cv}
\end{align}
where $\textbf{S} = \textbf{K}(\textbf{D},\textbf{D})[\textbf{K}(\textbf{D},\textbf{D}) + \zeta_0 \textbf{I}]^{-1}$. 

The posterior predictive mean of $f$ used as an emulator (in line 6) is,
\begin{equation}
\bar f = \textbf{K}(\textbf{d}_{\star}, \textbf{D})[\textbf{K}(\textbf{D},\textbf{D}) + \zeta_0 \textbf{I}]^{-1}\tilde{U}(\textbf{D}|\textbf{d}_c)
\end{equation}
for arbitrary inputs $\textbf{d}_{\star}$ derived using standard results on the conditional distribution of normal random variables. 
Finally, computing the design efficiency (line 7) is achieved by normalising the functional utility to
\begin{equation}
eff(\textbf{d}_{\star j}) = \ffrac{\bar f(\textbf{d}_{\star j}|\textbf{D},\hat{\boldsymbol{\zeta}})}{\bar f(\hat{\textbf{d}}_{*j}|\textbf{D},\hat{\boldsymbol{\zeta}})}
\label{eqn:eff}
\end{equation}
where $\hat{\textbf{d}}_{\star j} = \underset{\textbf{d}_{\star j} \in \textbf{D}_{\star}}{\mathrm{arg\ max}}\ \bar f(\textbf{d}_{\star j}|\textbf{D},\hat{\boldsymbol{\zeta}})$. Output design efficiency contours (line 8) as the set of prediction points above threshold value $t$.

\LinesNumbered
\begin{algorithm}
\SetAlgoLined

If existing, let $\textbf{d}_c = (\textbf{d}_{c1}, \textbf{d}_{c2},...,\textbf{d}_{c\gamma})$ be the current design. Specify $\tilde{U}(\textbf{d}|\textbf{d}_c)$ as appropriate and define prior $p(\boldsymbol{\theta})$ and likelihood $p(\textbf{y}|\boldsymbol{\theta}, \textbf{d})$. \\

\nonl\emph{Phase I - Construct $q$-dimensional utility grid} \\

Define design space $\mathcal{D} \subseteq \mathbb{R}^q$ across $q$ windows, sample $\textbf{D} = (\textbf{d}_1, ..., \textbf{d}_{\gamma_1})$ with $\textbf{d}_i \in \mathcal{D}$ for training and $\textbf{D}_{\star} = (\textbf{d}_{\star 1}, ..., \textbf{d}_{\star \gamma_2})$ with $\textbf{d}_{\star j} \in \mathcal{D}$ for predictions.

Compute median utility $\Tilde{U}(\textbf{d}_i)$ for each point $\textbf{d}_i$\\

\nonl\emph{Phase II - Functional utility surface} \\

Fit a GP; hyperparameters are determined by minimising cross-validation error\\
$\hat{\boldsymbol{\zeta}} = \underset{\boldsymbol{\zeta}}{\mathrm{arg\ min}}\ CV(\boldsymbol{\zeta}|f,\textbf{d})$ with CV defined in Equation \eqref{eqn:cv}

Evaluate predicted mean values $\bar f$ at each prediction point $\textbf{d}_{\star j}$ as
$\bar f(\textbf{d}_{\star j}|\textbf{D},\hat{\boldsymbol{\zeta}}) = \sum\limits^{\gamma_1}_{i=1} k(\textbf{d}_{\star j}, \textbf{d}_i| \hat{\boldsymbol{\zeta}})\alpha_i $ for $\boldsymbol{\alpha} = [\textbf{K}(\textbf{D},\textbf{D}) + \hat\zeta_0 \textbf{I}]^{-1} \tilde{U}(\textbf{D}|\textbf{d}_c)$\\

\nonl\emph{Phase III - Design Efficiency contours} \\

Normalise utility surface to yield efficiencies
$eff(\textbf{d}_{\star j})$ as per Equation \eqref{eqn:eff} for each prediction point $\textbf{d}_{\star j}$

Output set of points where $eff(\textbf{d}_{\star}) > t$ for threshold values $t = [0,1]$

\caption{Sampling Windows Algorithm}\label{gp}
\end{algorithm}

\section{Case Studies} \label{sec:casestudy}


\subsection{Example 1: River Network Design}

\begin{figure}
\includegraphics[width=1\textwidth]{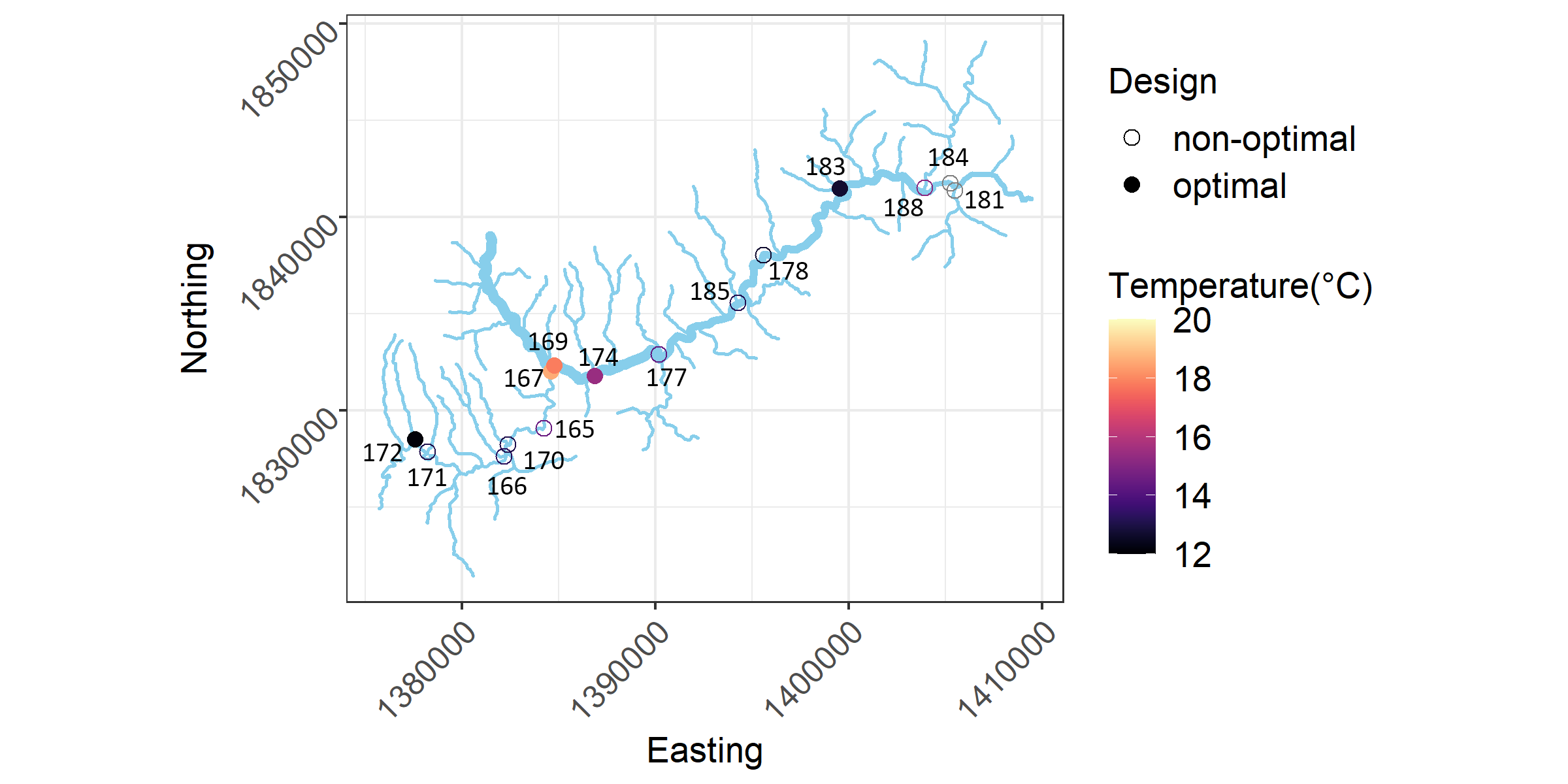}
\caption{A total of 15 observation sites located in the Clearwater River Basin, USA. Average daily water temperature for July 1, 2013. The width of the stream line is proportional to Shreve’s stream order \citep{shreve1966statistical}. The optimal and non-optimal design locations using the proposed stochastic Coordinate Exchange search are represented by filled and unfilled dots, respectively.}
\label{fig:clearwater}
\end{figure}

The design problem in this case study is motivated by a costly and complex balance of priorities, that is, short-term interest in monitoring particular river segments (e.g. wildfire impacts on juvenile fish habitat), while maintaining the ability monitor efficiently at established locations throughout the region.
In this example, we use a water temperature dataset from the Clearwater River Basin, USA, with observation sites located from 500m up to 46km apart (Figure \ref{fig:clearwater}). 
Temperature data were collected at 15 spatial locations using in-situ sensors that recorded measurements at 30 minute intervals \citep{isaak2018principal}. 
The response variable is mean temperature for July 1, 2013, with stream slope, elevation, watershed area, and air temperature used as covariates \citep{santos2022ssnbayes}.
Note that spatial variation in temperature often constitutes the largest proportion of total variation, since temporal correlation among sites is strong \citep{isaak2017norwest}.
A tail-down covariance function was found to be the most suitable covariance function for describing this spatial dependence.



There are 15 existing monitoring sites in the Clearwater River Basin and the design problem is to find the best 5 among them.
In addition, wildfires occurred near two segments that provide cold water refugia for juvenile fish and sensors are to be deployed to monitor impacts on water temperature. 
For convenience, we define search neighbourhoods along the river network, that is, any continuous line along the branching river network and specify Neighbourhood 1 (N1) and Neighbourhood 2 (N2) in Figure \ref{fig:windowsnetwork}.

\begin{figure}
\includegraphics[width=0.5\textwidth]{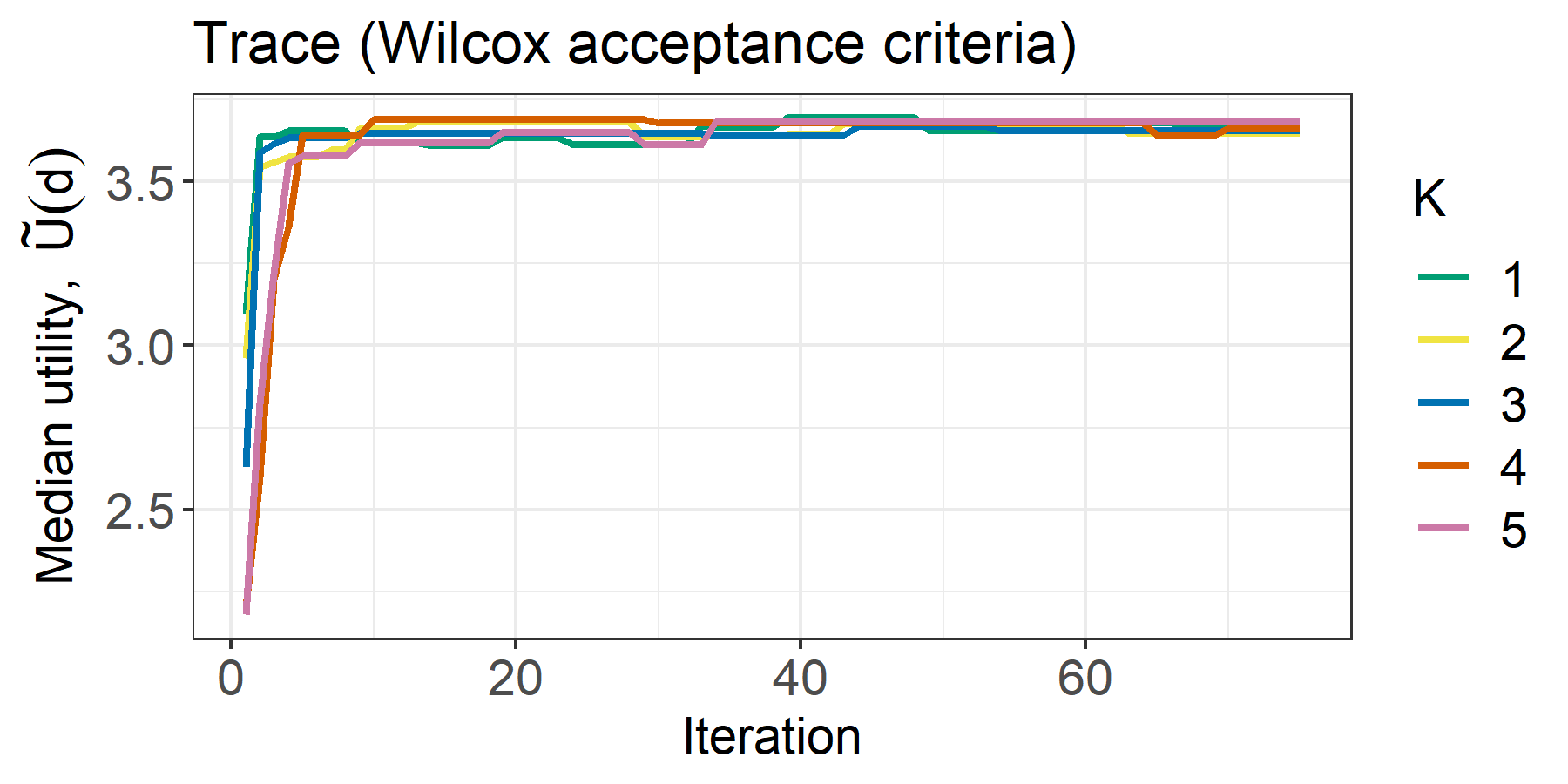}
\includegraphics[width=0.5\textwidth]{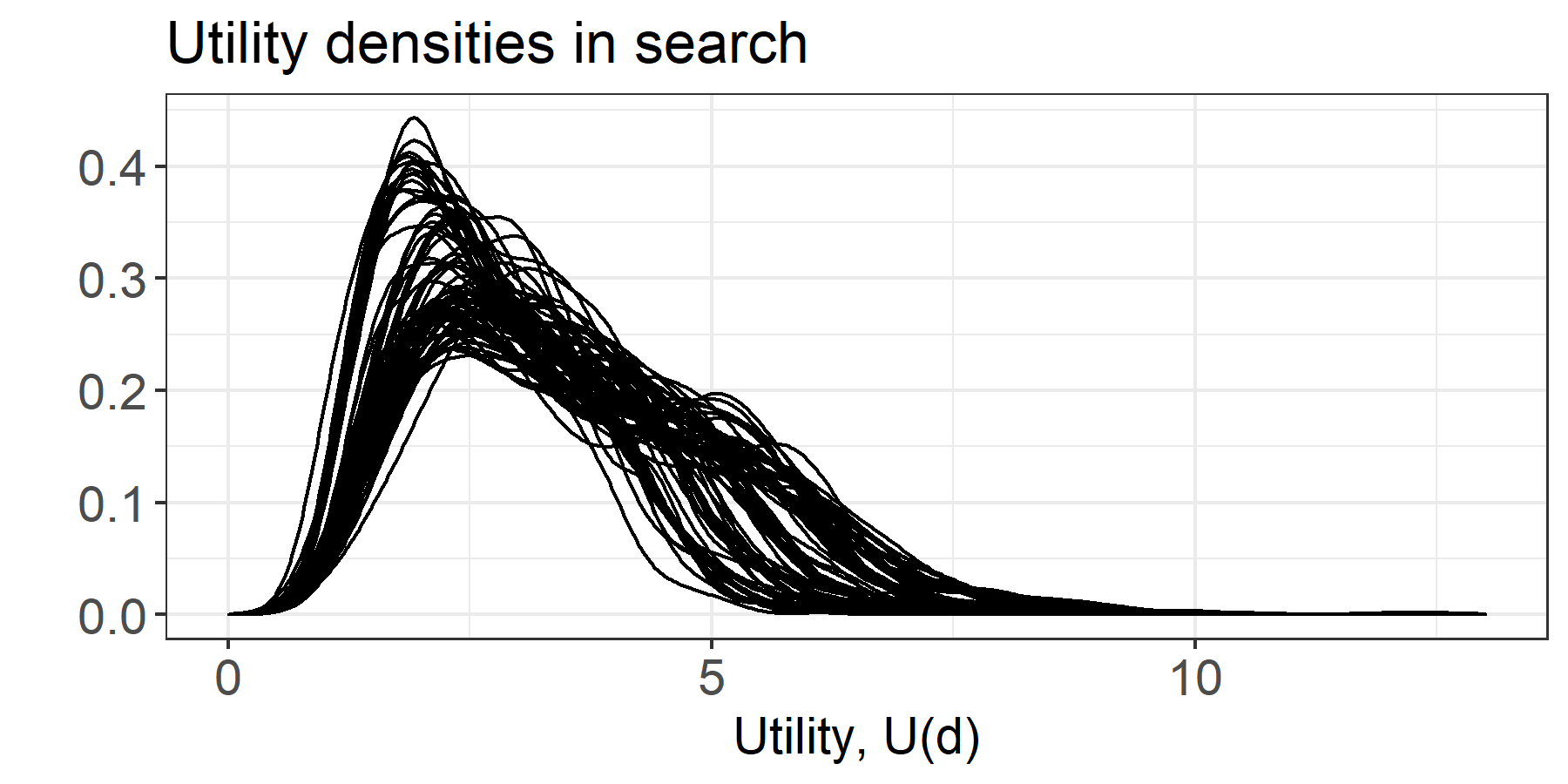}
\includegraphics[width=0.5\textwidth]{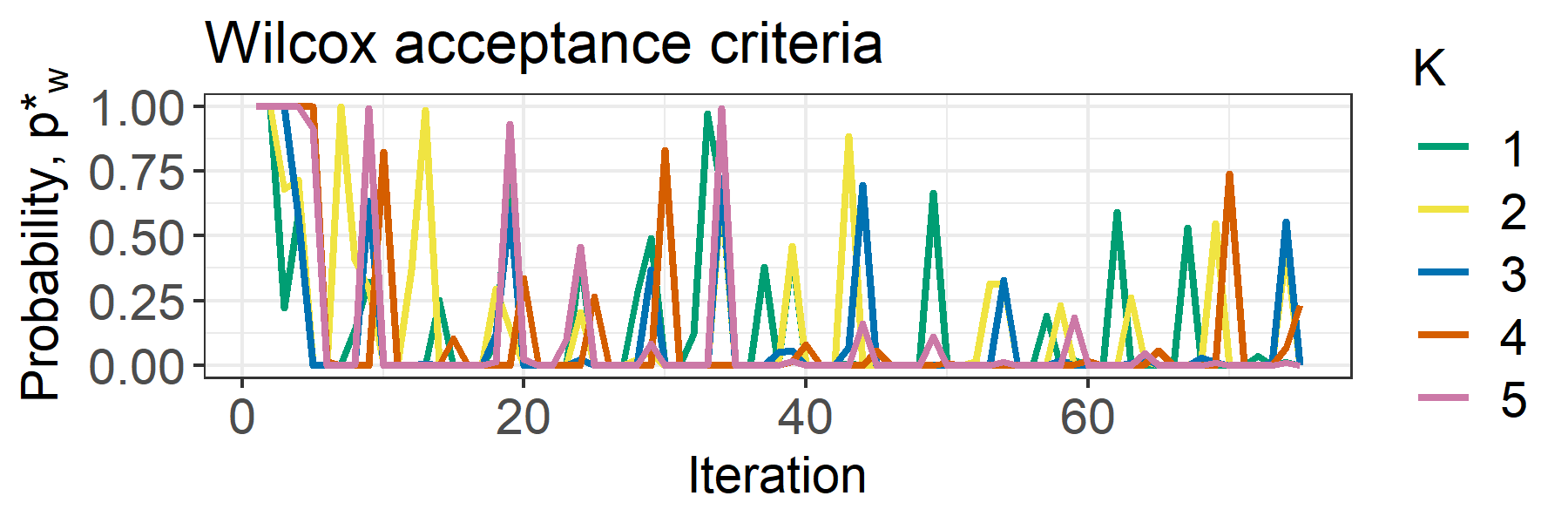}
\includegraphics[width=0.5\textwidth]{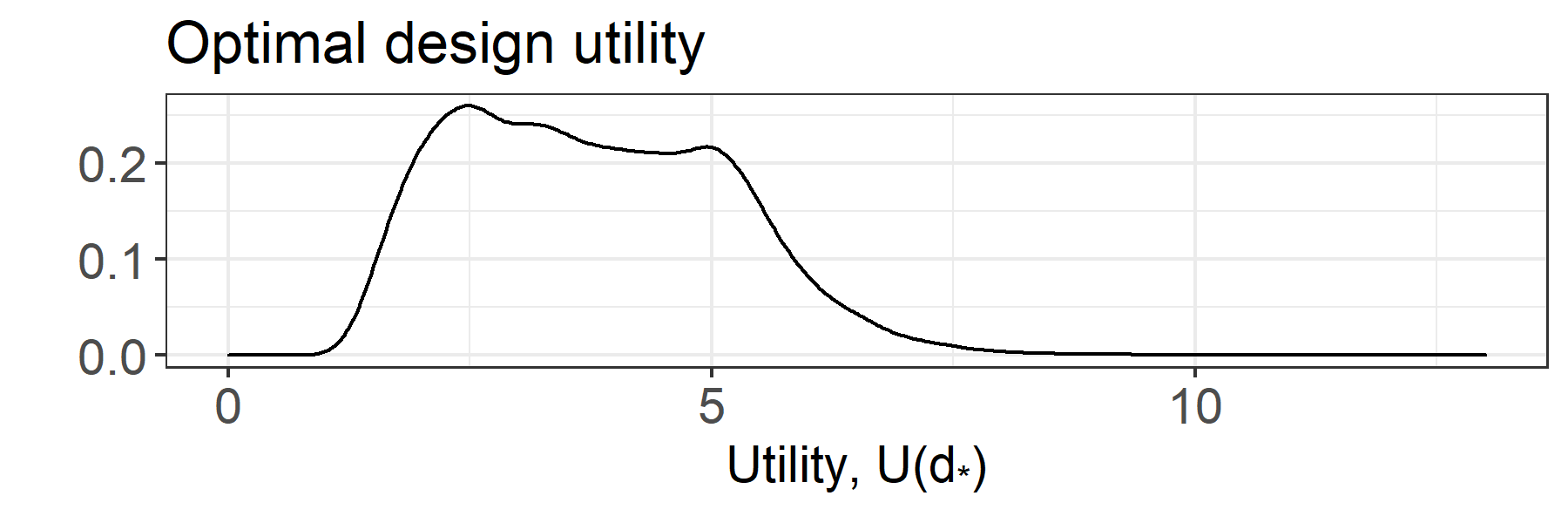}
\label{figutility}
\caption{Trace plot for the coordinate exchange algorithm using Wilcoxon acceptance criteria and median utility (top-left); proposal probability using Wilcoxon acceptance criteria, $p^*_W$ (bottom-left); every 50th utility distribution in the proposed CE algorithm (top-right); and the optimal utility distribution (bottom-right).}
\end{figure}

The stochastic Coordinate Exchange approach outlined in Algorithm \ref{stochastic_search} (Appendix) was computed with $K = 5$ random starts and $T=15$ outer iterations. 
Figure \ref{figutility} shows a selection of the utility distributions found in the proposed search algorithm, then the distributions of the optimal design, noting the non-normality of these distributions. 
The optimal design was found to be at locations, $\textbf{d}_* = (167, 169, 172, 174, 183)$ with a median utility of 3.6805. 
These optimal locations cover a wide range for each of the covariate values, in particular, optimal locations include the maximum and minimum value for air temperature. 
The set of locations $(167, 169, 174)$ are clustered at a river confluence and the most downstream location of all sites, $169$, is included in the design (Figure \ref{fig:clearwater}). 
These characteristics agree with optimal designs found using previous pseudo-Bayesian approaches for river networks \cite{falk2014sampling}. 
We also benchmark the performance of the proposed Coordinate Exchange algorithm against other variations; a comparison of acceptance criteria (Wilcoxon vs. ACE), as well as central tendency measure (mean vs. median) is provided in the Appendix. 
The choice of the median as a preferred measure is supported by the apparent skewness and long tails of the distributions of the utilities (Figure \ref{figutility}).

Recall the specification of neighbourhoods N1 and N2 within the river network for sensor deployment.
The proposed Sampling Windows Algorithm (Algorithm \ref{gp}), is applied.
Consider the design space $\mathcal{D} \subseteq \mathbb{R}^2$ with design points $d_i^1 \in N1$ and $d_i^2 \in N2$ for each design $\textbf{d}_i = (d_i^1, d_i^2)$.
The current design $\textbf{d}_c$ is set as the optimal points $\textbf{d}_*$ from the downsizing regime previously found.
In Phase I, the median utility is computed as $\tilde{U}(\textbf{d}_i | \textbf{d}_c) = \tilde{U}(\textbf{d}_i \cup \textbf{d}_c)$ for each $\textbf{d}_i$ in the grid, $\textbf{D}$, sampled from neighbourhoods 500m apart.
Covariate values at $\textbf{d}_i$ for air temperature are interpolated using 3 nearest neighbours based on Euclidean distance, whereas the remaining covariate values (stream slope, elevation and watershed area) are unique to each river segment. 
In Phase II, a GP is fit assuming a kernel function defined by Equation \eqref{eq:kernel}, where the distance measure $\mbox{dist}_j(\textbf{d}_s, \textbf{d}_p) = |d_s^j - d_p^j|$ for $j=1,2$ is equivalent to stream distance (along dimension $j$). 
Results from Phase III indicate that N1 is more sensitive to location placement than N2, as shown in Figure \ref{fig:windows}. 

From an applied standpoint, these design outputs allow decisions to be made in the field (e.g. placement around specific location $n_{15}$ in N1, but anywhere convenient on the main branch in N2). 
This provides support for greater flexibility in sensor placement when access issues arise, without having to rerun a full discrete optimal design search.
A practical approach, given the design efficiency contours as a roadmap, would be to visit N1 first; should $n_{15}$ be inaccessible then move along the river to deploy the in-situ sensor where possible, say $n_{14}$. 
Take the line segment corresponding to $eff(\textbf{D}_*|n_{14})$ to inform deployment in N2, in this case, $n_{24}$, to retain $98\%$ optimal utility.


\begin{figure}
\includegraphics[width=1\textwidth]{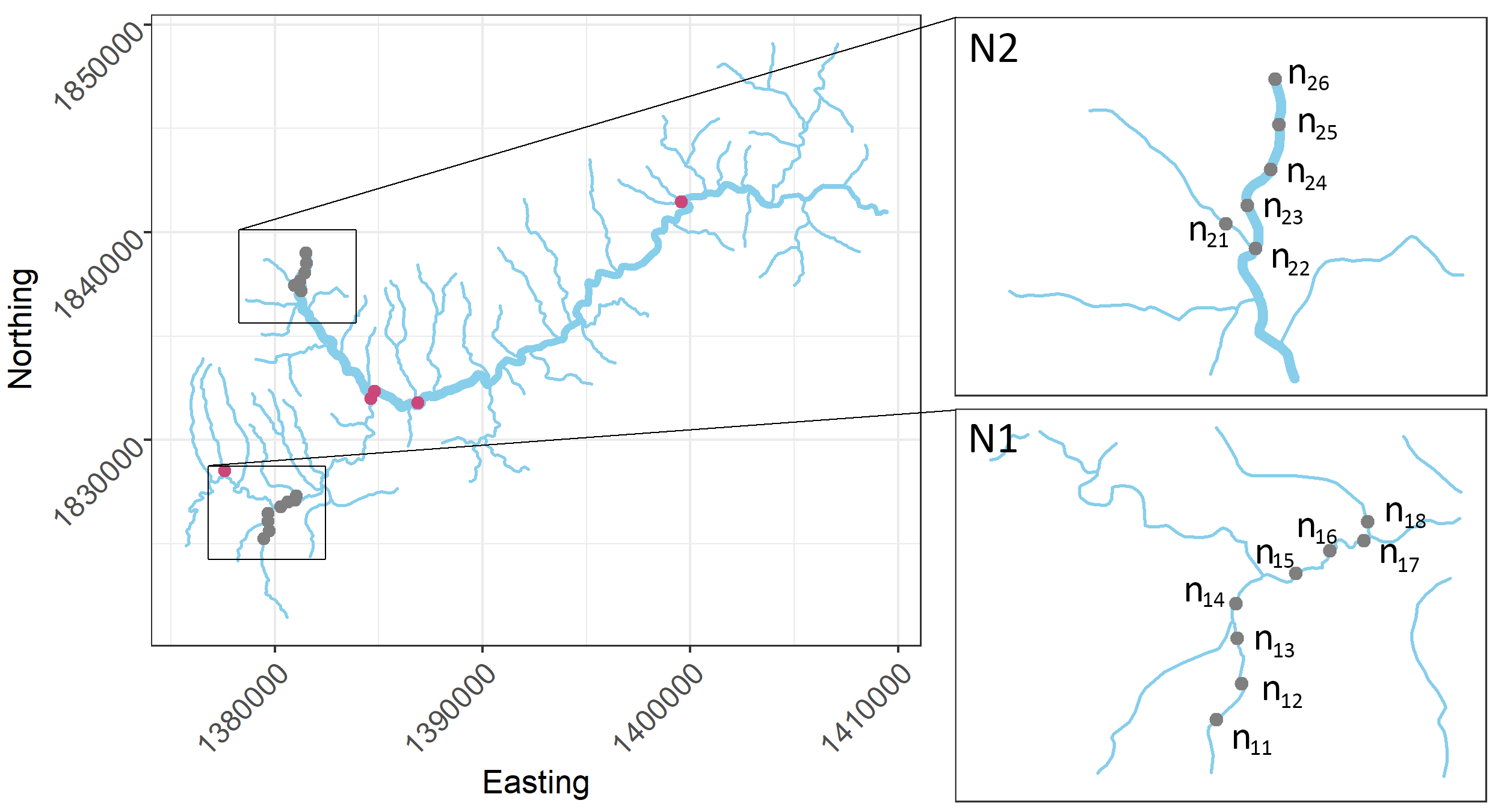}
\caption{Existing sites (red dots) and potential sites within search neighbourhoods N1 and N2 (gray dots) located on the river network. A discrete set of points were sampled from both neighbourhoods spaced at 500m stream distance apart.}
\label{fig:windowsnetwork}
\end{figure}

\begin{figure}
\includegraphics[width=1\textwidth]{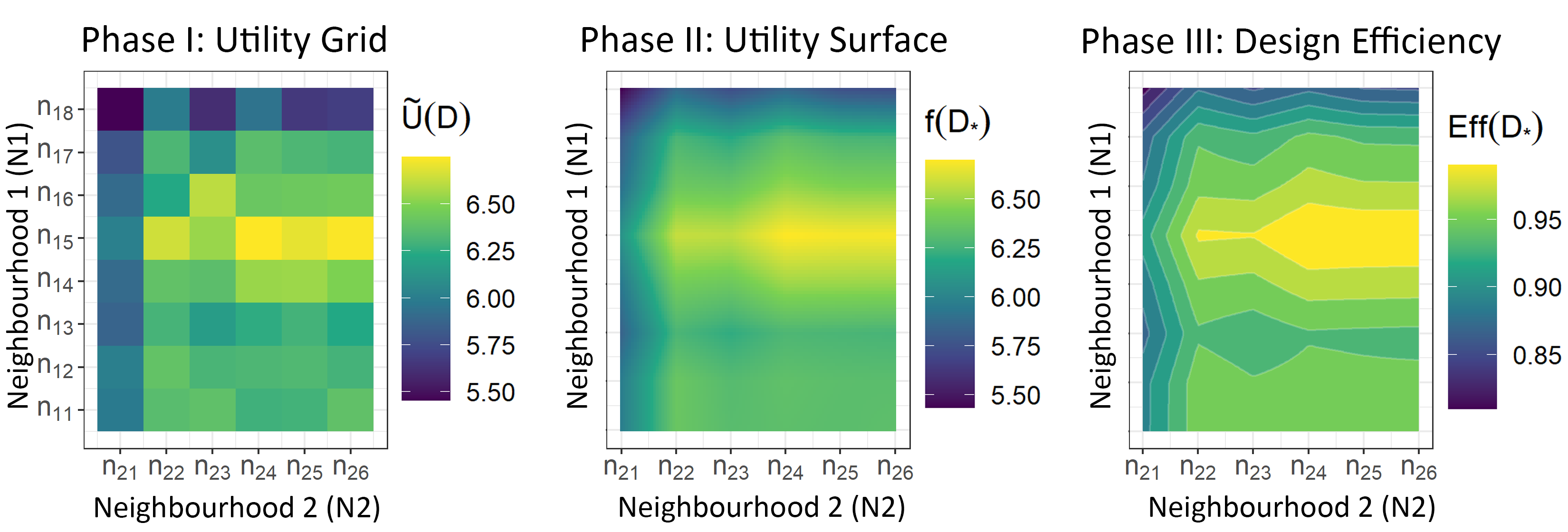}
\caption{Phases of the proposed Sampling Windows Algorithm; constructed median utility grid (left), Gaussian process mean predictions (middle), and normalised contours of design efficiency (right). A sensor close to $n_{15}\in$N1 will yield high design efficiency, while more flexibility can be afforded for the choice of sensor placement in N2.}
\label{fig:windows}
\end{figure}

\subsection{Coral Reef Design}

The Australian Institute of Marine Science (AIMS) has monitored the condition of coral reefs around Australia for decades.
We considered data collected in 2013 from the Barracouta East submerged coral reef off the north-west coast of Australia \citep{heyward2013montara}.
The design problem is to find the best regions to sample underwater images in. 
We propose a two step approach using the methods described above: first, a global search for the optimal transect lines with $n=3$, the total number of transects, using the proposed stochastic CE algorithm in a discrete design space; then, a continuous search for the optimal radius of each transect region with $q=3$, the number of sampling windows centred at each transect.

Coral cover is assessed using images of the reef floor taken every 5m along a transect, where $n_k = 20$ random points on image $k$ are analysed, leading to the consideration of the logistic regression model described in Equation \eqref{eqn:logit}.
Exploratory analyses showed a strong linear relationship between the response variable, $\textbf{q}$, and depth of the reef floor which was used as the only covariate. 
Note that depth was interpolated using 3 nearest neighbours based on Euclidean distance from the survey data. 
To reduce the design space and thus the complexity of the search, angles are discretised to $\alpha^j = \{0, 45, 90, 135\}$, $E^j$ and $N^j$ are defined by a discrete set of points. Length remains fixed at $l^j=500$ so that each transect compromises of 100 images.

The proposed Stochastic Coordinate Exchange algorithm with the Wilcoxon acceptance criteria and median utility (Algorithm \ref{stochastic_search}) was applied to find the optimal transect lines across the shoal.
When computing the utility, the Laplace approximation to the posterior requires evaluating the full data likelihood, as shown in Equation \eqref{eqn:fulllikelihood}. 
Accordingly, the spatially correlated random effect, $\textbf{z}$, needs to be integrated out. 
For this, we chose to use Monte Carlo integration with $M_z=50$ draws from the marginal likelihood due to computational restraints.
The Monte Carlo integration for the median utility in Equation \eqref{eq:monte} is evaluated with $M=350$ and $B=600$ draws from the prior and full likelihood for the evaluation of $\hat{U}(\textbf{d}^{j\curvearrowright})$ and $p^*_W$, respectively.

\begin{figure}
\includegraphics[width=0.48\textwidth]{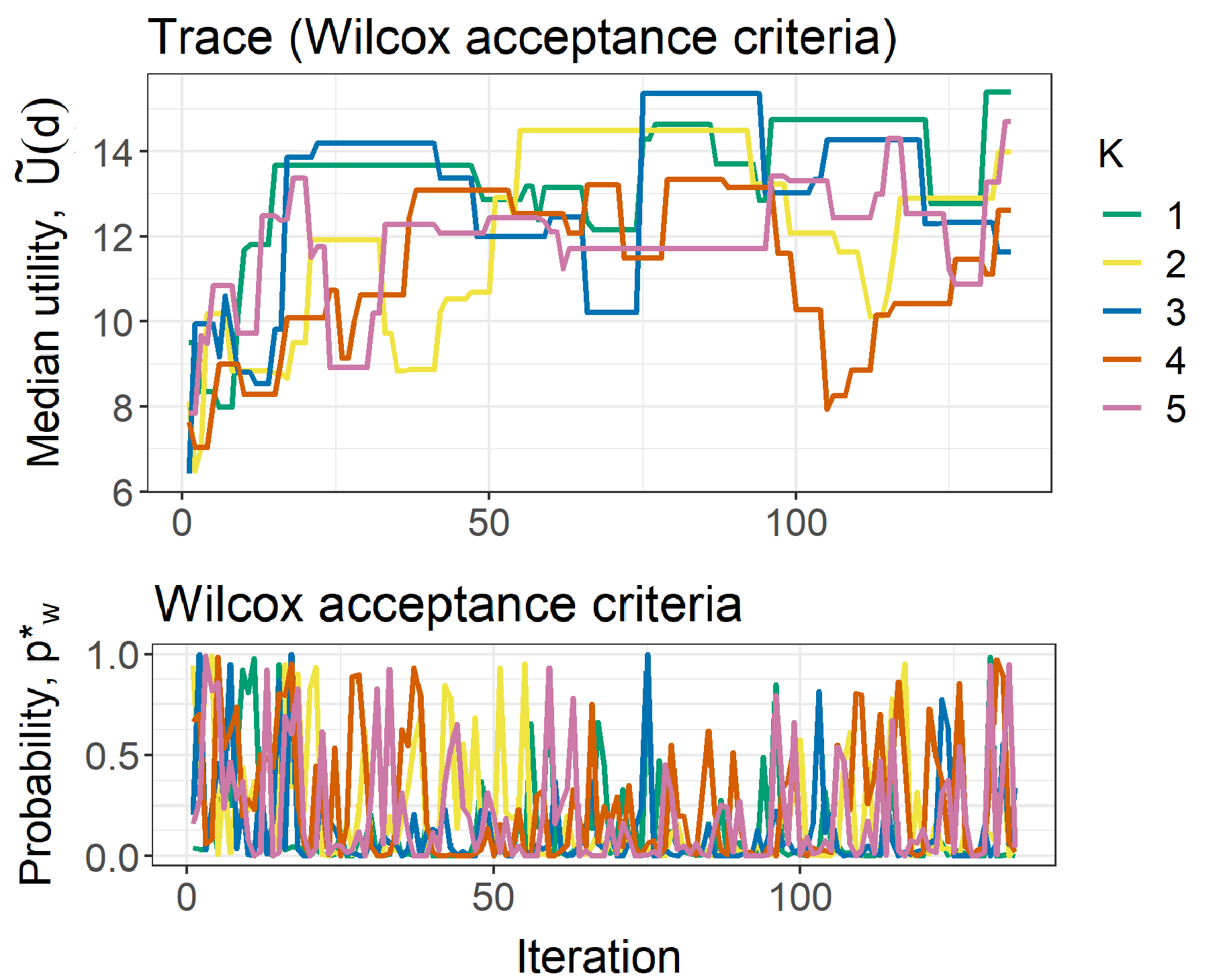}
\includegraphics[width=0.52\textwidth]{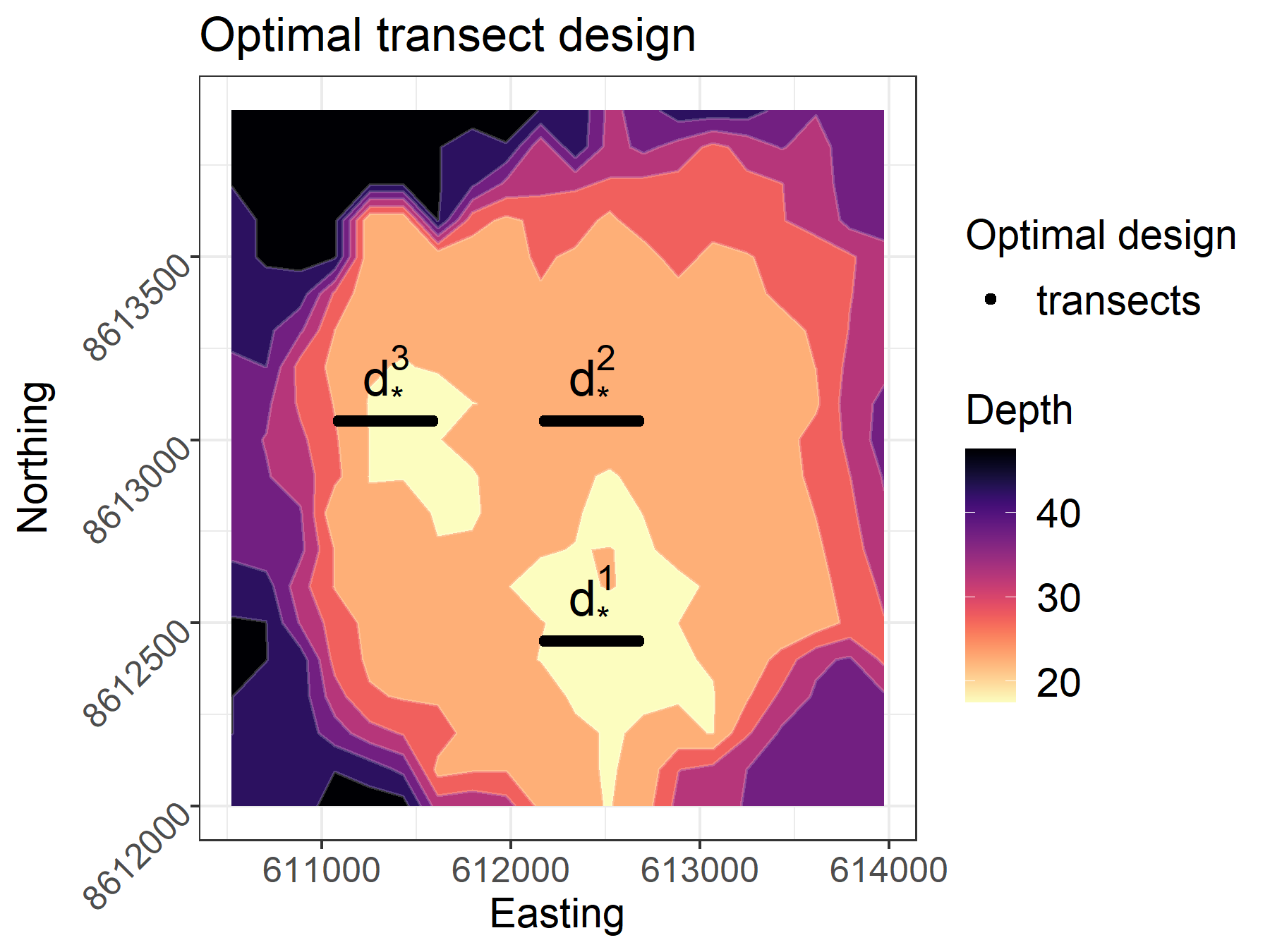}
\caption{Trace plot of proposed stochastic coordinate exchange algorithm (top-left), acceptance probability (bottom-left), and optimal transect design in geographic space (right).}
\label{fig:optimal_transect}
\end{figure}

Spatial covariance is characterised by exponential decay, defined in Equation \eqref{eq:exp}.
To further ease the computational burden, we construct a spatial grid (10 $\times$ 10) over the entire shoal, with $h_{_{EUC}}$ the Euclidean distance between grid centre points (a lower-dimensional surrogate to the $N \times N$ distance matrix using coordinates $\textbf{x}_k$).
This grid size is a compromise; not so small that the computational gains are marginal, but not so large that the spatial relationships are not captured between grid rectangles. 
Results in Figure \ref{fig:optimal_transect} show the optimal placement of transects $\textbf{d}_* = (d_*^1, d_*^2, d_*^3)$ in the shallower areas of the reef, each with varying depth gradients. 
Thus, it would seem reasonable to sample in along these transects to estimate coral cover.

\begin{figure}
\includegraphics[width=\textwidth]{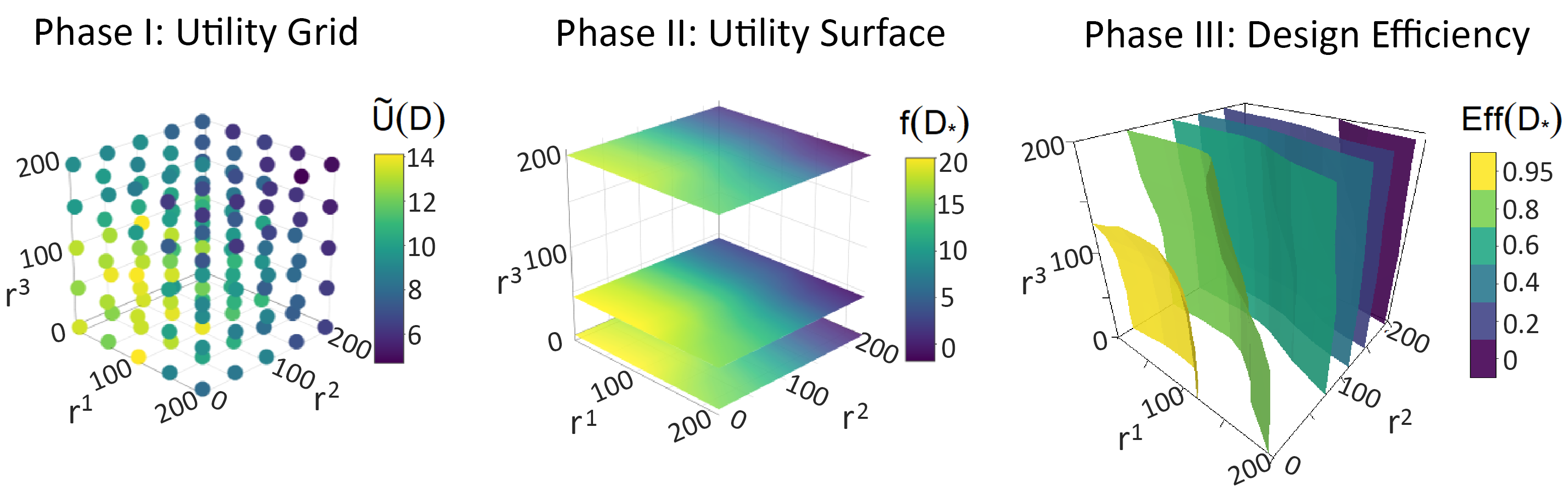}
\caption{The median utility (left); Gaussian process mean predictions of median utility (middle), and normalised contours of design efficiency (right) found at each phase of the proposed Sampling Windows Algorithm. The radius parameter, $r^j$, specifies a radius for transect $d^j_*$ for $j = \{1,2,3\}$.}
\label{3dwindows}
\end{figure}

The proposed Sampling Windows Algorithm (Algorithm \ref{gp}) is applied over the parameter space $r$, with the kernel function defined by Equation \eqref{eq:kernel} and distance $\mbox{dist}_j(r_s, r_p) = |r_s^j - r_p^j|$ for $j=1,2,3$. 
In the first step, the optimal transect design previously found is fixed, while the design space $\mathcal{D} \subseteq \mathbb{R}^3$ is constructed for radius parameter, $r$, corresponding to each transect.
Since stochasticity is introduced in the choice of geographic locations, $\hat{s}_k$, we average median utility across 3 location samples for a given radius, each with $M=600$ Monte Carlo draws from the prior and likelihood.
The 3-dimensional design contours are shown in Figure \ref{3dwindows}.
Optimal design efficiency, $eff(\textbf{D}_*) = 101\%$, is achieved at $r_* = (44, 0, 44)$.
This implies optimality when sampling within a region for the shallower areas, and more strictly across the transect $d_*^2$ in the deeper area of the reef.
Generally, these contours can be used to guide decision making on the field, for example, sample $d_*^2$ when conditions are good, whereas sampling in $d_*^1$ and $d_*^3$ can be done with stronger underwater currents.

\section{Discussion}

Coral reefs and freshwater ecosystems are biodiversity hotspots that are particularly sensitive to anthropogenic impacts of climate change, water pollution and over-exploitation, among other factors, resulting in rapid population declines \citep{strayer2010freshwater, tickner2020bending, wagner2020coral}. 
Informative data are needed to support critical and timely conservation decisions, but various practical constraints make precise sampling difficult or even impossible. 
In this paper, we have considered Bayesian optimal design for collecting data where greater flexibility is needed in the design outputs. 
To this end, we proposed a spatial Sampling Windows Algorithm using Gaussian process emulation to estimate design efficiency contours. 
While we focused on two important environmental systems, such an approach is also applicable to a broad range of complex systems that require rigorous design. 
For example, optimal design methods for sub-sampling can be used as a computationally efficient way to analyse Big Data \citep{drovandi2017principles}. Rather than extracting optimal design points from the Big Data set, which may or may not exist, points could be selected within windows.

There are several novel aspects to the approaches described here.
Statistical models for branching river networks use distance travelled along the river network, directionality, and flow volume to describe spatial relationships. 
There is novelty in the application of Bayesian design for river network systems, as of yet to be presented in the statistical literature.
Since utility distributions were not normally distributed, we also proposed the use of a non-parametric acceptance probability in the stochastic Coordinate Exchange, based on the one-sided Wilcoxon Rank Sum Test.
We successfully demonstrated how Bayesian design works in complex spatial settings and introduced a method to find sampling windows, facilitating intelligent data collection.

The novel methodological advances presented here provide practical solutions to common monitoring challenges.
Results from the proposed Sampling Windows Algorithm highlighted in the river network case study showed that design efficiency was best achieved around a smaller range in Neighbourhood 1, while almost agnostic to choice in Neighbourhood 2; recalling that neighbourhoods are any continuous line along the river.
In particular, the design efficiency contours identify which, and quantify how much, design points are sensitive to precise sampling.
Similarly, results in the coral reef case study showed high design efficiency around regions in shallow areas of the reef compared to deeper regions with less hard coral.
Providing fine-scale information about optimality allows practitioners in the field to make informed judgements on jointly optimal variables based on physical conditions not included in the utility (e.g. accessibility, safety, etc.), without the need to re-run time consuming optimisations.


There are also numerous opportunities for future research in this space. 
One useful extension would be to incorporate additional monitoring objectives related to cost into the utility function.
For example, the optimal deployment of a combination of high-cost sensors that produce high-quality data and lower-cost sensors that produce less reliable data.
Sampling and monitoring costs are a core consideration for any monitoring program, but it is especially true in remote locations. 
In such cases, it would also be useful to consider the travel time between sampling locations in the optimal design. 
New Bayesian spatio-temporal models for river networks \citep{santos2021bayesian} provide the opportunity to extend the approach developed here for spatio-temporal sampling windows. 
This would allow practitioners to make more effective use of mobile in-situ sensors for spatio-temporal data collection. 
Thus, the novel methods presented here (along with future extensions) facilitate more efficient data collection; providing insights into spatial (and spatio-temporal) processes which can be used to better inform data-enabled management decisions in complex and vulnerable ecosystems.

\section*{Acknowledgments}

This project was supported by the Australian Research Council (ARC) Linkage Project (LP180101151) ``Revolutionising water-quality monitoring in the information age". 
We thank Dan Isaak and Dona Horan from the US Forest Service Rocky Mountain Research Station for sharing the data and creating the SpatialStreamNetwork object used in the river network case study. 
We also acknowledge the Seascape Health and Resilience team of the Australian Institute of Marine Science (AIMS) for the curation and provision of data towards the coral reef case study.

\section*{Code Accessibility}

The computations were performed using the R package bayesDesign, which contains functions for fitting models, optimisations and visualising design domains. The code can be found online at \url{github.com/KatieBuc/bayesDesign}. Note that the code relies on high performance computing infrastructure.

\bibliographystyle{rss}
\bibliography{mybib}

\newpage 
\appendix
\renewcommand\thesection{\Alph{section}}
\renewcommand{\thefigure}{\thesection.\arabic{figure}}

\section{Appendix}

\subsection{Covariance on River Networks}

Again, two points $d^k$ and $d^s$ on a river network are said to be \emph{flow-connected} if they share water flow, and \emph{flow-unconnected} if they reside on the same network and do not share flow. 
The tail-up covariance between locations separated by stream distance $h_{_{H}}$ can then be defined as follows:

\begin{equation*}
C_{_{TU}}(h_{_{H}}| \boldsymbol{\theta}_{\textbf{z}_{TU}}) = 
	\begin{cases}
		\textbf{W}_{ks} C(h_{_{H}}|\boldsymbol{\theta}_{\textbf{z}_{TU}})  & \text{if flow-connected},\\
		0 &  \text{if flow-unconnected},
	\end{cases}
\end{equation*}

where $\textbf{W}_{ks}$ represents the spatial weights between $d^k$ and $d^s$ defined by the branching structure of the river network and $C(h_{_{H}}|\boldsymbol{\theta}_{\textbf{z}_{TU}})$ is the (unweighted) exponential covariance function defined in Equation \eqref{eq:exp}. 
The tail-down covariance between locations separated by stream distance $h$ is defined by the exponential covariance function $C_{_{TD}}(h_{_{H}}|\boldsymbol{\theta}_{\textbf{z}_{TD}})=C(h_{_{H}}|\boldsymbol{\theta}_{\textbf{z}_{TD}})$
noting that $h_{_{H}}$ is the hydrologic distance separating $d^k$ and $d^s$.
Similarly, the covariance between locations based on Euclidean distance is $C_{_{EUC}}(h_{_{EUC}}|\boldsymbol{\theta}_{\textbf{z}_{EUC}}) = C(h_{_{EUC}}|\boldsymbol{\theta}_{\textbf{z}_{EUC}})$.  Note that, the exponential model is the only known covariance function that produces a statistically valid covariance matrix when Euclidean distance is replaced with total hydrologic distance \citep{ver2006spatial}. 
In matrix notation, $\textbf{C}_{_{TU}}$, $\textbf{C}_{_{TD}}$ and $\textbf{C}_{_{EUC}}$ are matrices derived from the above covariance functions, respectively.

\subsection{Stochastic Coordinate Exchange Algorithm}

In Algorithm \ref{stochastic_search}, an initial random design $\textbf{d}_0 = (d_0^1, ..., d_0^\gamma)$ is chosen.
In the inner-most for loop, coordinate $j$ in the design $\textbf{d}$ is swapped for a point in the design space $\mathcal{D} \setminus \textbf{d}$, denoted by $\textbf{d}^{j\curvearrowright}$.
With each swap, the median utility is computed. 
Then, the best found utility $\tilde{U}(\textbf{d}^{j\curvearrowright})$ is compared to the current best utility $U_{best}$ via the acceptance criteria. 
If accepted, the best design is updated. 
There are $K$ random starts to mitigate convergence to a local maxima, which can be computed in (embarrassing parallel) fashion. 

\begin{algorithm}[H]
Specify $\tilde{U}(\textbf{d})$ as appropriate. Define prior $p(\boldsymbol{\theta})$ and likelihood $p(\textbf{y}|\boldsymbol{\theta}, \textbf{d})$. Define design space, $\mathcal{D}$, as set of points with $\Gamma = |\mathcal{D}|$.\\

\SetAlgoLined
\For{$k=1:K$}{
    Initial random design, $\textbf{d}_0$\\
    Set $\textbf{d}_k = \textbf{d}_0$ and evaluate $U_{best} = \tilde{U}(\textbf{d}_k)$\\
    \For{$t=1:T$}{
        \For{$j = 1:\gamma$}{
            \For{$\iota = 1:(\Gamma-\gamma)$}{
            Set $\textbf{d}_k^{j\curvearrowright}$ as $\textbf{d}_k$ with the $j$th coordinate exchanged with the $\iota$th design point not in $\textbf{d}_k$\\
            Evaluate $\tilde{U}(\textbf{d}_k^{j\curvearrowright})$, with large $M$
            }
        Acceptance probability $p_W^* = 1- W(\mbox{max}(\tilde{U}(\textbf{d}_k^{j\curvearrowright})), U_{best})$, with larger $B$\\
        If accepted, $\textbf{d}_k = \mbox{arg max }\tilde{U}(\textbf{d}_k^{j\curvearrowright})$ and $U_{best} = \mbox{max}(\tilde{U}(\textbf{d}_k^{j\curvearrowright}))$\\
        }
    }
}
Evaluate $\tilde{U}(\textbf{d}_k)$ with largest $B_{k}$, $\textbf{d}_* = \mbox{arg max }\tilde{U}(\textbf{d}_k)$
 \caption{Stochastic Coordinate Exchange Algorithm}\label{stochastic_search}
\end{algorithm}

\begin{figure}
\includegraphics[width=0.5\textwidth]{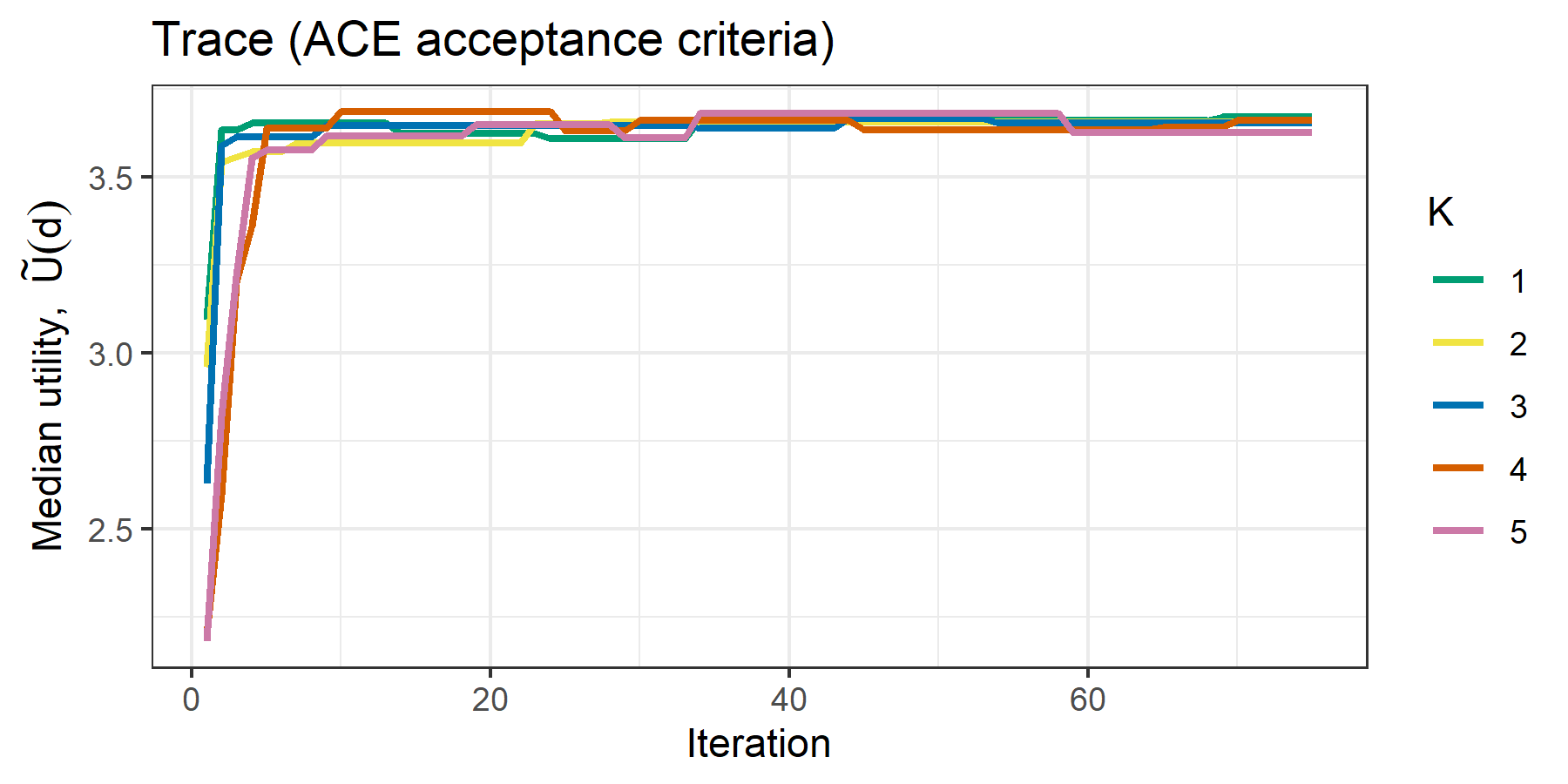}
\includegraphics[width=0.5\textwidth]{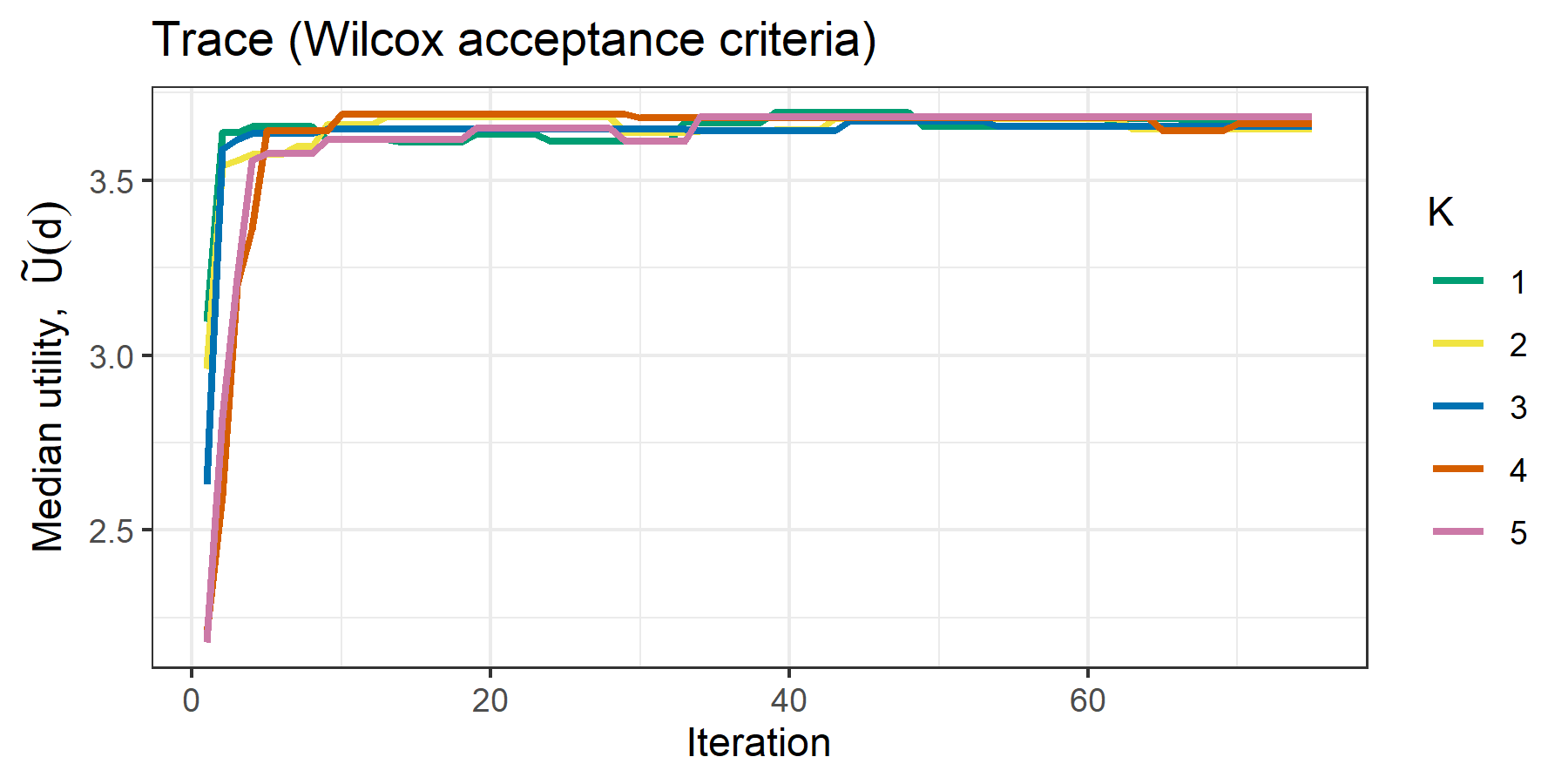}
\includegraphics[width=0.5\textwidth]{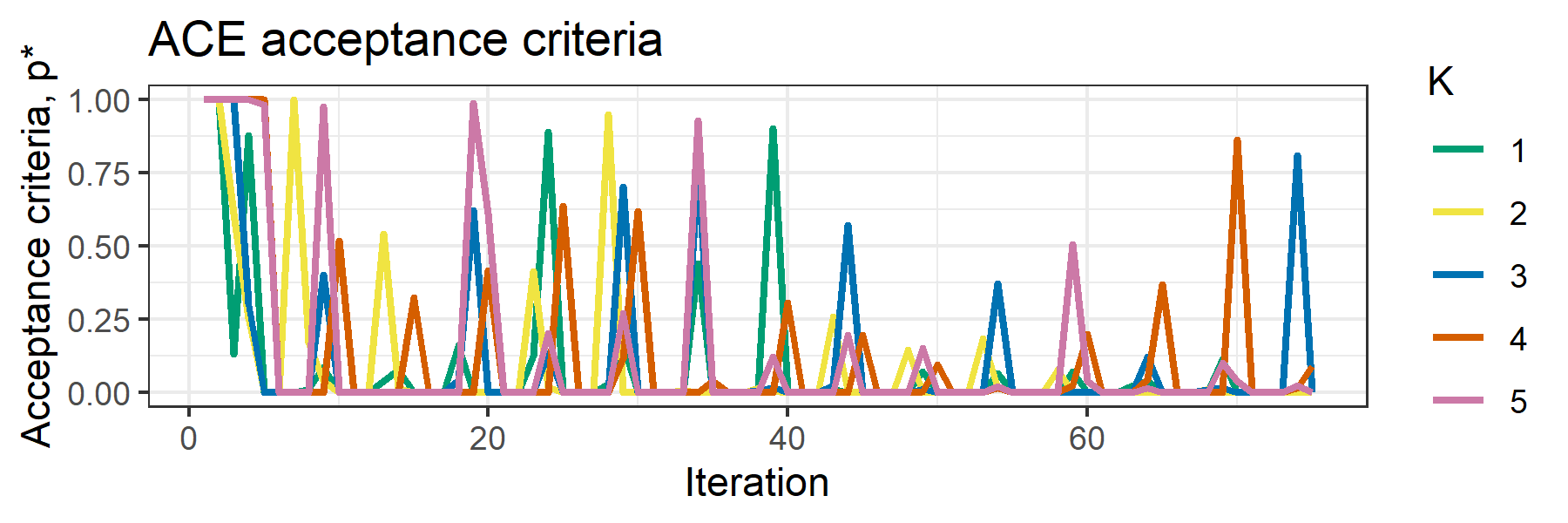}
\includegraphics[width=0.5\textwidth]{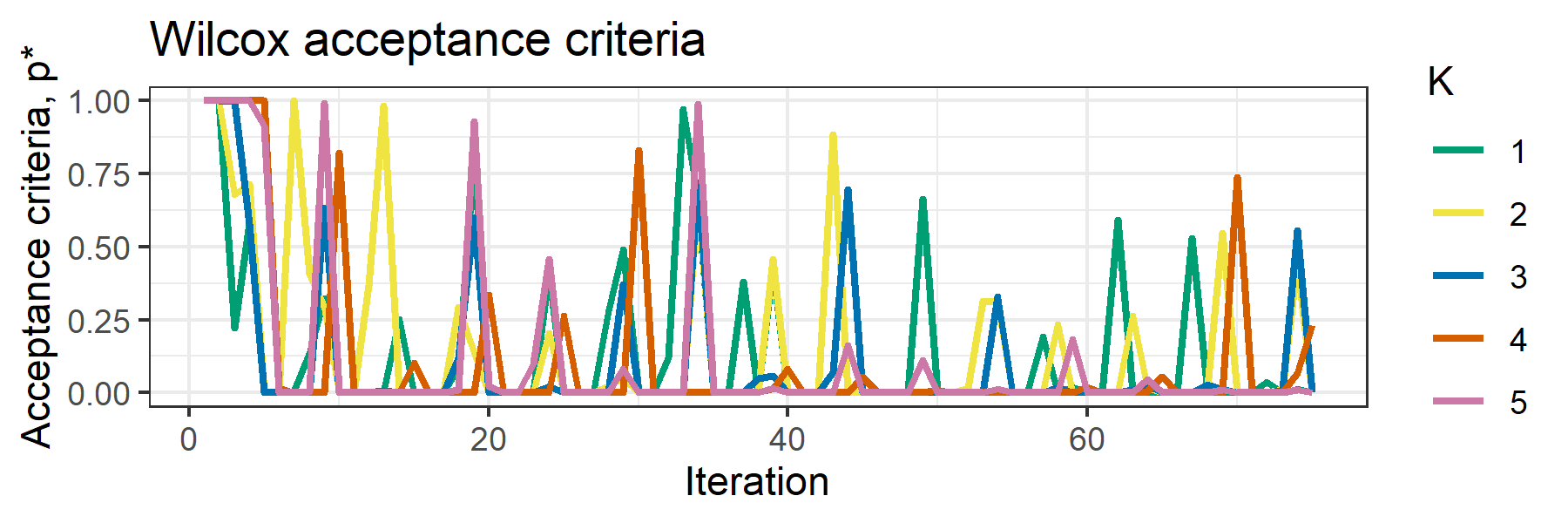}
\includegraphics[width=0.5\textwidth]{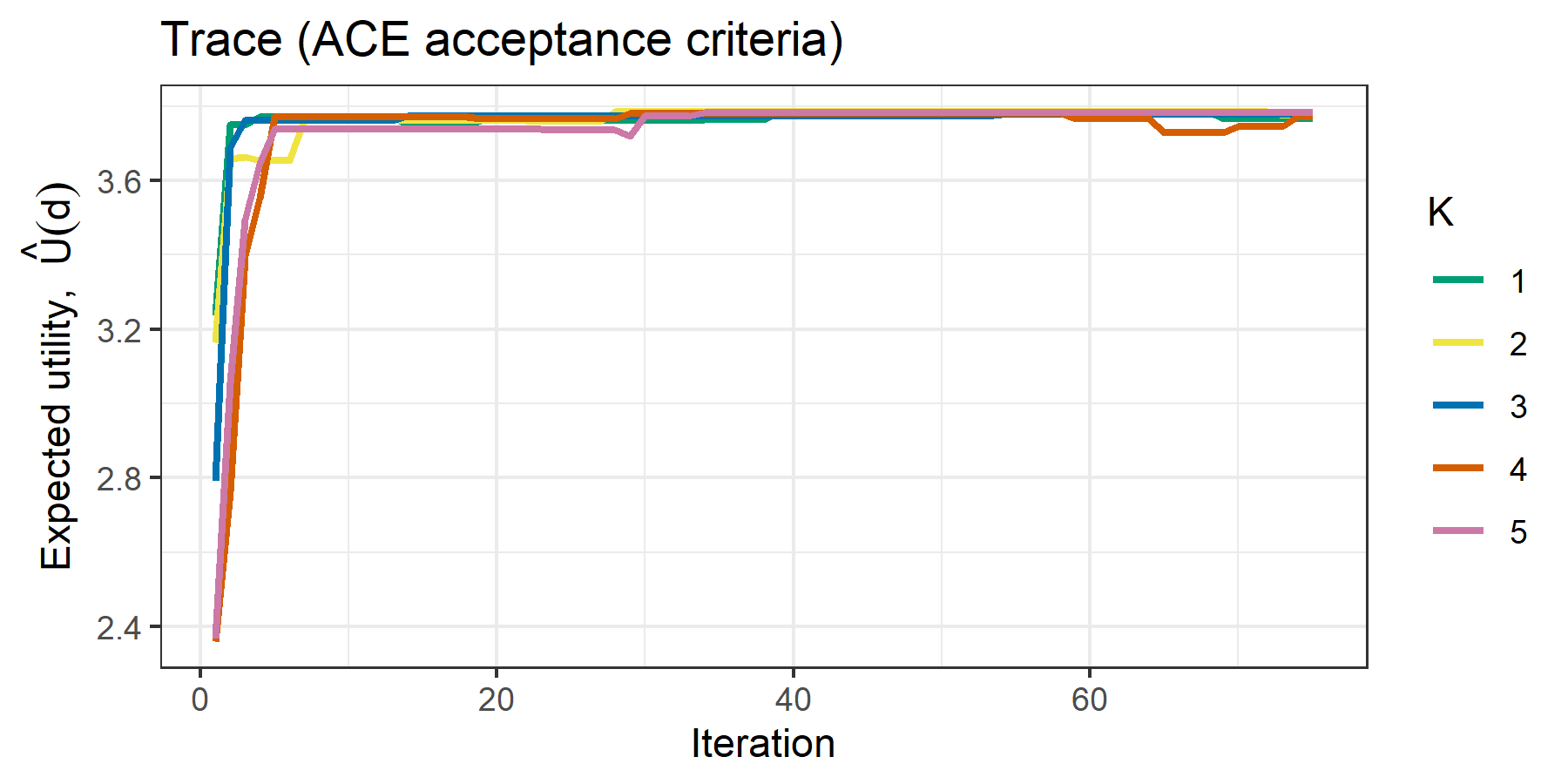}
\includegraphics[width=0.5\textwidth]{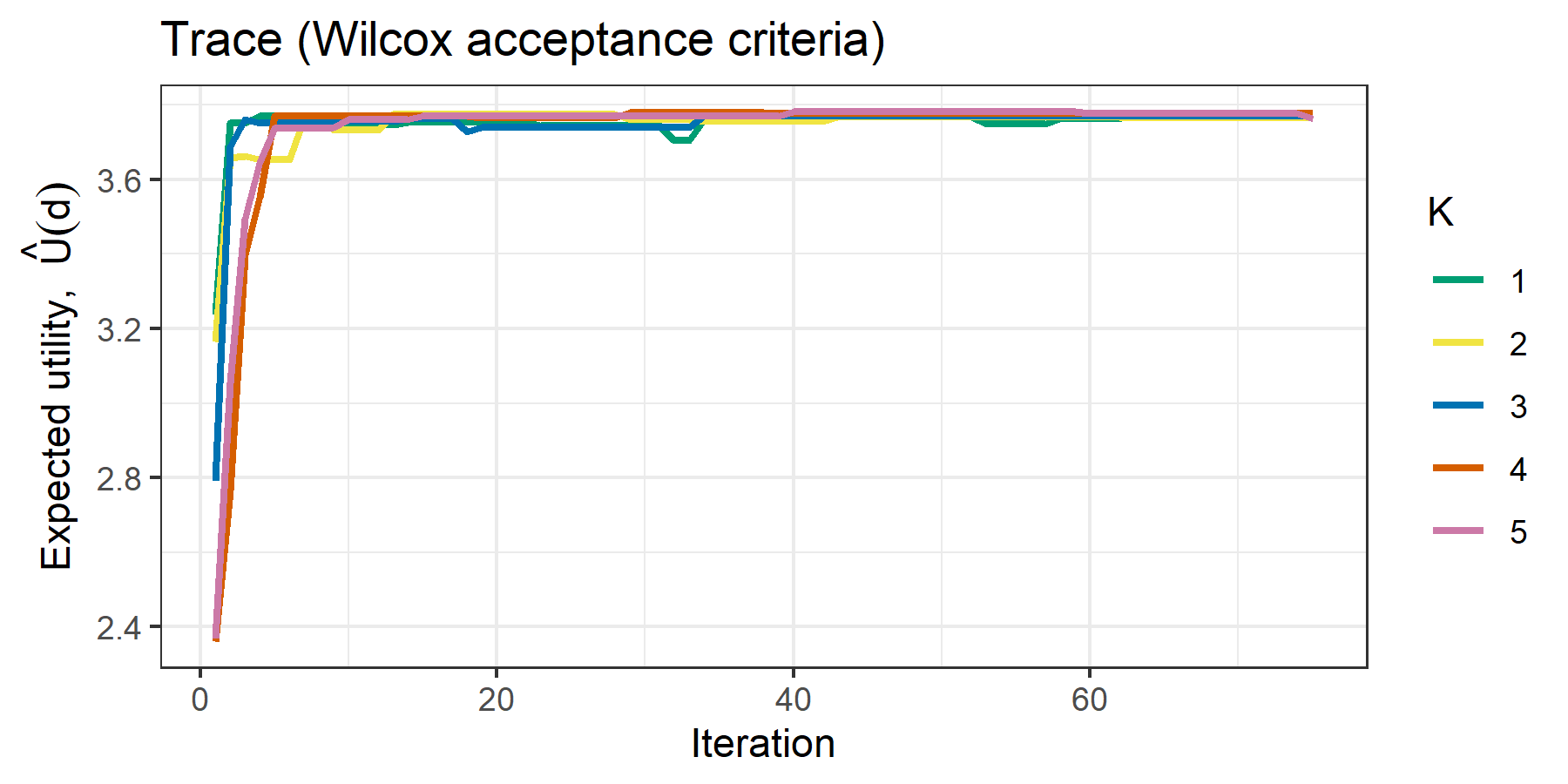}
\includegraphics[width=0.5\textwidth]{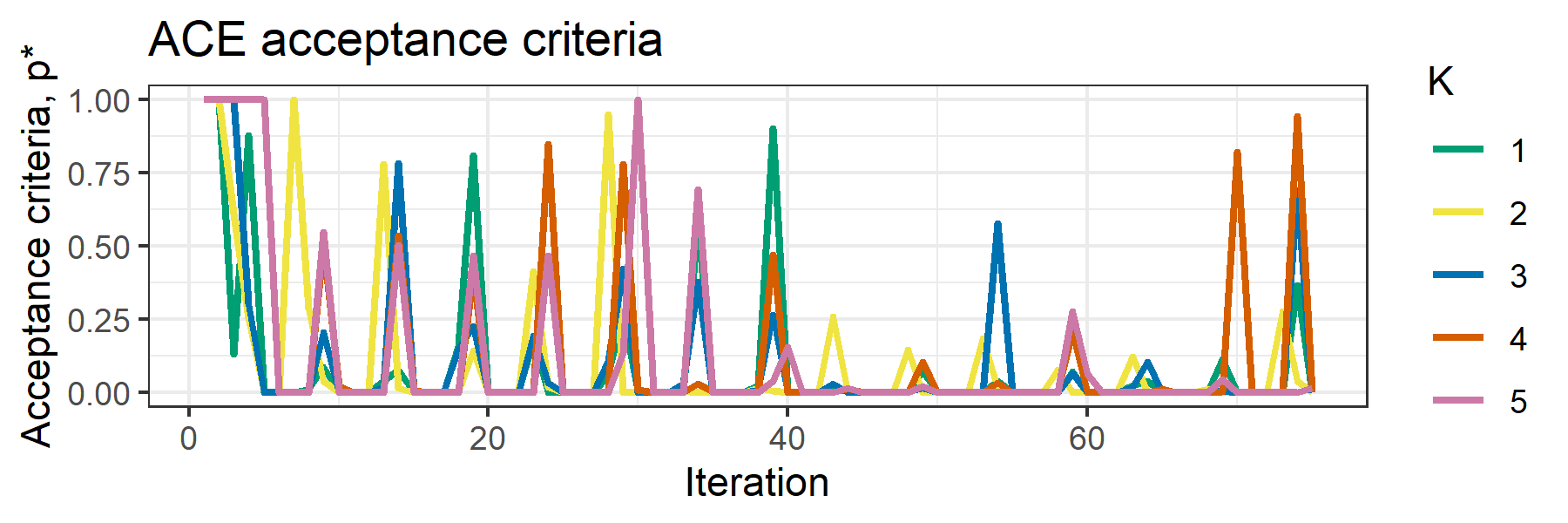}
\includegraphics[width=0.5\textwidth]{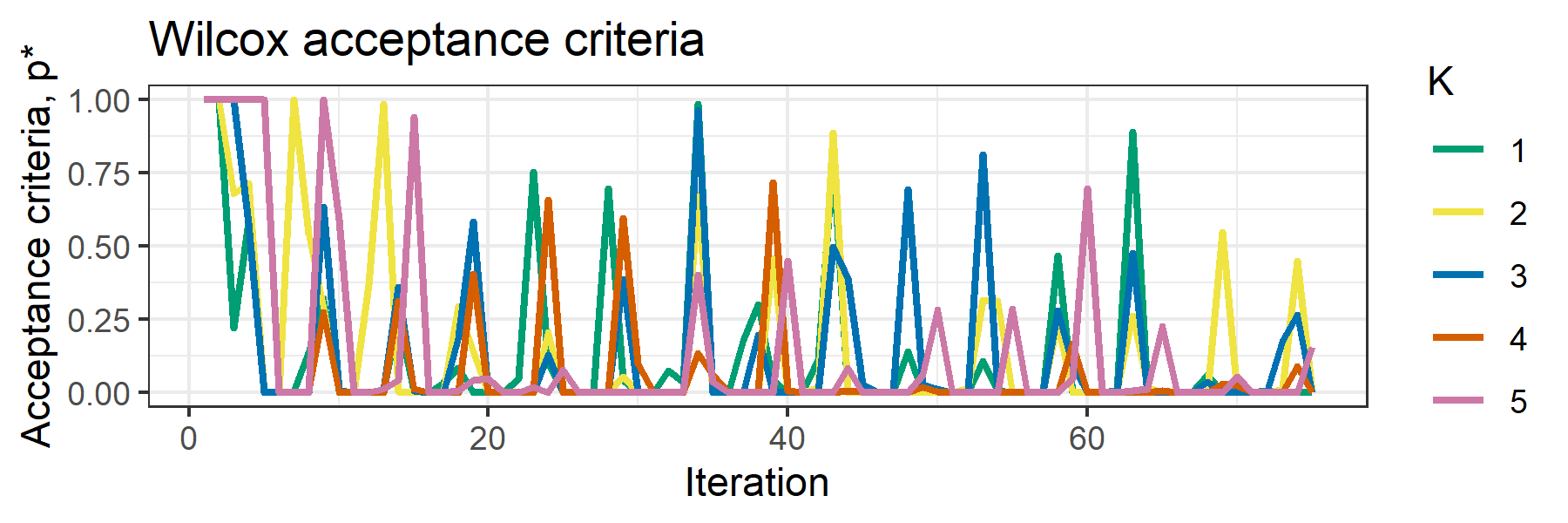}
\caption{Stochastic Coordinate Exchange Algorithm trace plots using ACE acceptance criteria (left) and the proposed Wilcoxon acceptance criteria (right) for the expected KL divergence utility.}
\label{fig:benchmark}
\end{figure}

\begin{table}
\caption{Optimal design scenarios for the river network case study}
\begin{tabular}{lll}
Acceptance criteria & Utility measure & Optimal design\\ \hline
Wilcoxon & Mean  & (167, 169, 172, 174, 183)\\
ACE & Mean & (167, 169, 172, 174, 183)\\ \hline
Wilcoxon & Median & (167, 169, 172, 174, 183)\\
ACE & Median & (167, 169, 172, 174, 183)\\ \hline
\end{tabular}
\label{optimal_results}
\end{table}

In our example, the Wilcoxon and ACE acceptance criteria resulted in the same optimal design, with respect to both the mean and median utility CE algorithm (Table \ref{optimal_results}).
It is noted that with more uncertain priors or multi-modal utilities, the comparative results between acceptance criteria may be more pronounced.
Both optimal designs, and indeed most of the best designs for each random start, contain the points at confluence (167, 169, 174).

\subsection{Median Utility: MCMC vs. Laplace}

In general, the form of the posterior is unknown, and therefore difficult to estimate directly. 
In such cases, algorithms like Markov Chain Monte Carlo (MCMC) (e.g. Metropolis–Hastings algorithm) are employed to draw samples which are used to approximate the posterior distribution \citep{hastings1970monte}.

To efficiently approximate the expected utility in Bayesian design, fast methods for approximating the posterior distribution are required. 
For this, we consider the Laplace approximation to the posterior distribution, see \cite{long2013fast, overstall2018approach, carlon2020nesterov, senarathne2020bayesian}.
There are some advantages in using the Laplace approximate method compared to numerical Markov chain Monte Carlo (MCMC) methods. 
Besides the huge gains in computational efficiency for low-dimensional design problems, there is no issue of choosing a suitable proposal density, for which to tune for a desired acceptance rate, and similarly no need to ensure convergence of a Markov chain, as well as no need to determine the length of a suitable ``burn-in" phase. 
However, the Laplace approximation is limited to the assumption of an approximately normal target distribution, has been shown to underestimate posterior variance \citep{shun1997another}.

\begin{figure}
\begin{center}
\includegraphics[width=0.65\textwidth]{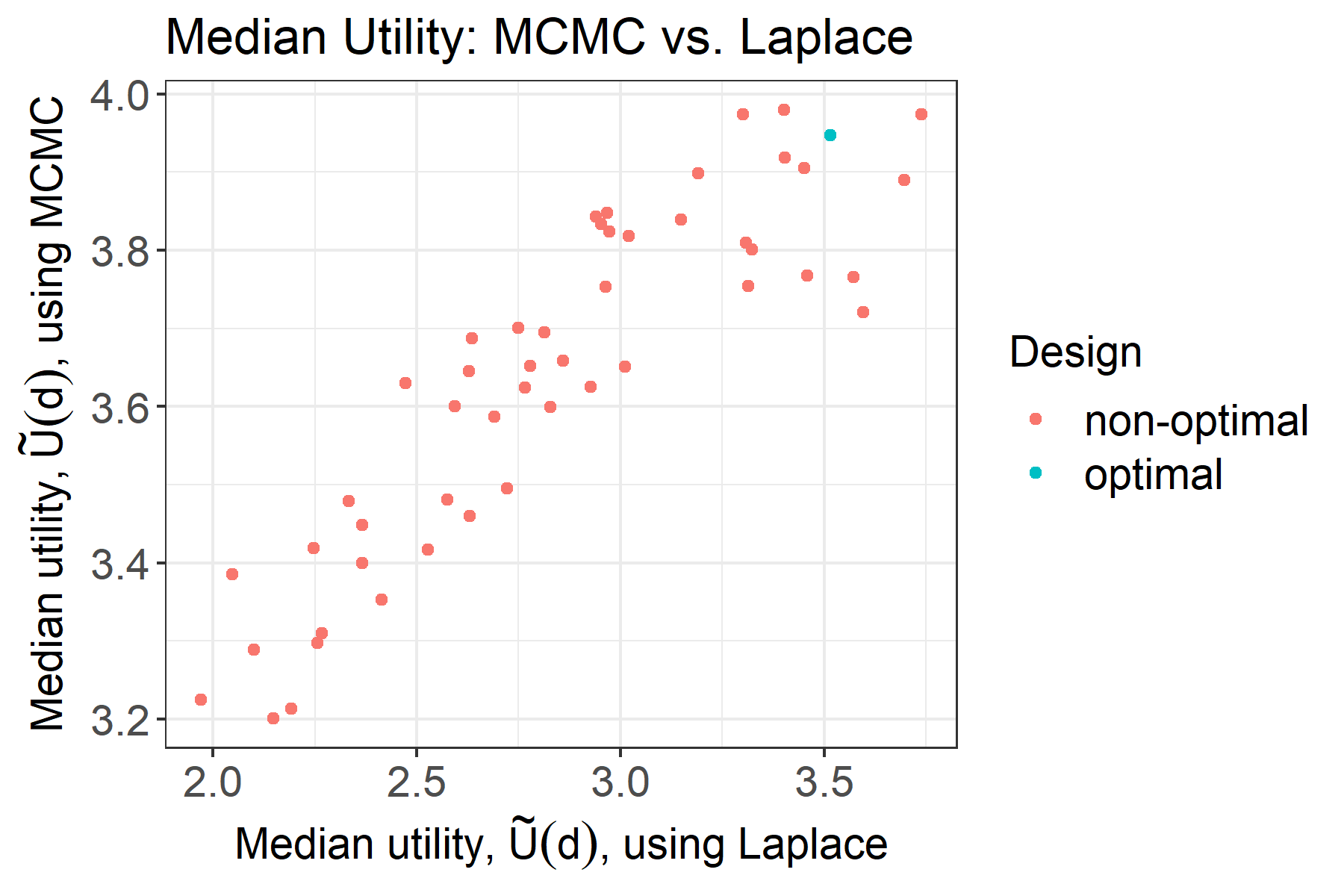}
\end{center}
\caption{The median utility, $\tilde U(\textbf{d})$, computed for 50 random designs using inference methods: MCMC (Metropolis–Hastings algorithm) vs. Laplace approximation. The optimal design found via the proposed stochastic Coordinate Exchange algorithm (Algorithm \ref{stochastic_search}) is shown in blue.}
\label{fig:mcmc}
\end{figure}

Therefore, we compared the median utility, $\tilde U(\textbf{d})$, using the Laplace approximation vs. an MCMC (Metropolis–Hastings algorithm) approximation to the posterior for 50 random designs. 
Figure \ref{fig:mcmc} shows the linear relationship between the Laplace and MCMC inference methods for the KL-divergence utility in the design problem outlined in Section 5.1.

\end{document}